\documentclass[sigplan, acmsmall, screen]{acmart}

\usepackage[framemethod=TikZ]{mdframed}
\usepackage{lipsum}
\mdfdefinestyle{promptFrame}{%
    linecolor=blue,
    outerlinewidth=2pt,
    roundcorner=20pt,
    innertopmargin=\baselineskip,
    innerbottommargin=\baselineskip,
    innerrightmargin=20pt,
    innerleftmargin=20pt,
    backgroundcolor=gray!10!white}
\usepackage{tabularx}
\usepackage{tcolorbox}
\usepackage{enumitem}
\usepackage{algorithmic}
\usepackage{graphicx}
\usepackage{textcomp}
\usepackage{booktabs} 
\usepackage{threeparttable}
\usepackage{framed}
\usepackage{verbatim}
\usepackage{pifont}
\usepackage{listings}
\newcommand{\xmark}{\ding{55}}%
\usepackage{xurl} 
\usepackage{tablefootnote}
\usepackage{xcolor}
\usepackage{hyperref}
\usepackage{xspace}
\usepackage{tikz}
\usepackage[T1]{fontenc}
\usepackage{beramono}
\hypersetup{
    colorlinks,
    linkcolor={red!50!black},
    citecolor={blue!50!black},
    urlcolor={blue!80!black}
}
\usepackage{filecontents}
\def\BibTeX{{\rm B\kern-.05em{\sc i\kern-.025em b}\kern-.08em
    T\kern-.1667em\lower.7ex\hbox{E}\kern-.125emX}}

\usepackage{color, colortbl}
\usepackage{soul}
\usepackage{multirow}
\usepackage{subcaption}

\DeclareFixedFont{\ttb}{T1}{txtt}{bx}{n}{11} 
\DeclareFixedFont{\ttm}{T1}{txtt}{m}{n}{11}  

\definecolor{deepblue}{rgb}{0,0,0.5}
\definecolor{deepred}{rgb}{0.6,0,0}
\definecolor{deepgreen}{rgb}{0,0.5,0}

\definecolor{dkgreen}{rgb}{0,0.6,0}
\definecolor{gray}{rgb}{0.5,0.5,0.5}
\definecolor{GrayMed}{rgb}{0.7,0.7,0.7}
\definecolor{GrayLt}{rgb}{0.9,0.9,0.9}
\definecolor{palegreen}{rgb}{0.92, 1, 0.9}
\definecolor{paleblue}{rgb}{0.92, 0.90, 1}
\definecolor{mauve}{rgb}{0.58,0,0.82}

\lstdefinestyle{PythonInline}{
  language=python,
  showstringspaces=false,
  basicstyle={\small\ttfamily\color{black}},
  keywordstyle=\color{blue},
  commentstyle=\color{mauve},
  stringstyle=\color{dkgreen}
}

\newcommand{\pyinline}[1]{\lstinline[style=PythonInline]!#1!}

\lstnewenvironment{Python}{
  \lstset{frame=tb,
  language=python,
  aboveskip=1mm,
  numbersep=0pt,
  belowskip=3mm,
  showstringspaces=false,
  columns=flexible,
  basicstyle={\scriptsize\fontfamily{zi4}\selectfont\color{black}},
  numbers=left,
  numberstyle=\tiny\color{mauve},
  keywordstyle=\color{blue},
  commentstyle=\color{mauve},
  stringstyle=\color{dkgreen},
  breaklines=true,
  breakatwhitespace=true,
  tabsize=3}
}{}

\makeatletter
\newenvironment{btHighlight}[1][]
{\begingroup\tikzset{bt@Highlight@par/.style={#1}}\begin{lrbox}{\@tempboxa}}
{\end{lrbox}\bt@HL@box[bt@Highlight@par]{\@tempboxa}\endgroup}

\newcommand\btHL[1][]{%
  \begin{btHighlight}[#1]\bgroup\aftergroup\bt@HL@endenv%
}
\def\bt@HL@endenv{%
  \end{btHighlight}%
  \egroup
}
\newcommand{\bt@HL@box}[2][]{%
  \tikz[#1]{%
    \pgfpathrectangle{\pgfpoint{1pt}{0pt}}{\pgfpoint{\wd #2}{\ht #2}}%
    \pgfusepath{use as bounding box}%
    \node[anchor=base west, fill=orange!30,outer sep=0pt,inner xsep=1pt, inner ysep=0pt, rounded corners=3pt, minimum height=\ht\strutbox+1pt,#1]{\raisebox{1pt}{\strut}\strut\usebox{#2}};
  }%
}
\makeatother

\lstnewenvironment{Java}{
  \lstset{frame=tb,
  language=java,
  aboveskip=1mm,
  numbersep=0pt,
  belowskip=3mm,
  showstringspaces=false,
  columns=flexible,
  basicstyle={\scriptsize\fontfamily{zi4}\selectfont\color{black}},
  numbers=left,
  numberstyle=\tiny\color{mauve},
  keywordstyle=\color{blue},
  commentstyle=\color{mauve},
  stringstyle=\color{dkgreen},
  breaklines=true,
  breakatwhitespace=true,
  tabsize=3,
  moredelim=**[is][\btHL]{`}{`},
  moredelim=**[is][{\btHL[fill=green!30,draw=red,dashed,thin]}]{@}{@},}
}{}

\AtBeginDocument{%
  \providecommand\BibTeX{{%
    \normalfont B\kern-0.5em{\scshape i\kern-0.25em b}\kern-0.8em\TeX}}}

\setcopyright{acmcopyright}
\copyrightyear{2024}
\acmYear{2024}
\acmDOI{XXXXXXX.XXXXXXX}

\acmConference[FSE'24]{}{July 15--19,
  2024}{Porto de Galinhas, Brazil}
\acmPrice{15.00}
\acmISBN{978-1-4503-XXXX-X/18/06}
\newcommand{\ic}[1]{\begin{small}\texttt{#1}\end{small}}
\newcommand{\NLSpec}{\textit{nl2postcond}}
\newcommand{\GPTthree}{\ic{GPT-3.5}}
\newcommand{\GPTfour}{\ic{GPT-4}}
\newcommand{\opensource}{\ic{StarChat}}
\newcommand{\humanevalplus}{\textit{EvalPlus}}
\newcommand{\humaneval}{\textit{HumanEval}}
\newcommand{\dfj}{\textit{Defects4J}}
\usepackage{etoolbox}
\newbool{longVersion}
\setbool{longVersion}{false}

\newcommand{\rqa}{How well do LLM-generated postconditions formalize informal natural language intent?}
 
\newcommand{\rqb}{Can LLM-generated postconditions help catch real-world bugs?}
 
\newcommand{\Comment}[1]{}

\usepackage{tcolorbox}
\usepackage{graphicx}
\usepackage{subcaption}
\newtcolorbox{takeawaybox}{colback=blue!10,colbacktitle=blue!40,title=Takeaways,colframe=blue!40,coltitle=black}

\newtcbox{\inlinebox}[1][]{enhanced,
 box align=base,
 nobeforeafter,
 colback=blueish,
 size=small,
 left=0pt,
 right=0pt,
 boxsep=2pt,
 #1}

\begin{document}

\title[Can LLMs Transform NL Intent into Formal Method Postconditions?]{Can Large Language Models Transform Natural Language Intent into Formal Method Postconditions?}

\author{Madeline Endres}
\authornote{Work done while interning at Microsoft.}
\affiliation{%
  \institution{University of Michgain}
  \city{Ann Arbor}
  \state{MI}
  \country{USA}
}
\email{endremad@umich.edu}
\author{Sarah Fakhoury}
\affiliation{%
  \institution{Microsoft Research}
  \city{Redmond}
  \state{WA}
  \country{USA}
}
\email{sfakhoury@microsoft.com}
\author{Saikat Chakraborty}
\affiliation{%
  \institution{Microsoft Research}
  \city{Redmond}
  \state{WA}
  \country{USA}
}
\email{saikatc@microsoft.com}
\author{Shuvendu K. Lahiri}
\affiliation{%
  \institution{Microsoft Research}
  \city{Redmond}
  \state{WA}
  \country{USA}
}
\email{shuvendu@microsoft.com}

\renewcommand{\shortauthors}{M. Endres, S. Fakhoury, S. Chakraborty, S. Lahiri}
\begin{abstract}

Informal natural language that describes code functionality, such as code comments or function documentation, may contain substantial information about a program’s intent. However, there is typically no guarantee that a program's implementation and natural language documentation are aligned. In the case of a conflict, leveraging information in code-adjacent natural language has the potential to enhance fault localization, debugging, and code trustworthiness. In practice, however, this information is often underutilized due to the inherent ambiguity of natural language which makes natural language intent challenging to check programmatically. The ``emergent abilities'' of Large Language Models (LLMs) have the potential to facilitate the translation of natural language intent to programmatically checkable assertions. However, it is unclear if LLMs can correctly translate informal natural language specifications into formal specifications that match programmer intent. Additionally, it is unclear if such translation could be useful in practice. 

In this paper, we describe \NLSpec{}, the problem of leveraging LLMs for transforming informal natural language to formal method postconditions, expressed as program assertions. 
We introduce and validate metrics to measure and compare different \NLSpec{} approaches, using the correctness and {\it discriminative power} of generated postconditions. 
We then use qualitative and quantitative methods to assess the quality of \NLSpec{} postconditions, finding that they are generally correct and able to discriminate incorrect code. Finally, we find that \NLSpec{} via LLMs has the potential to be helpful in practice; \NLSpec{} generated postconditions were able to catch 64 real-world historical bugs from \dfj{}.
\end{abstract}

\ccsdesc[300]{General and reference~Metrics}
\ccsdesc[100]{General and reference~Validation}
\ccsdesc[300]{Software and its engineering~Correctness}
\ccsdesc[300]{Software and its engineering~Completeness}
\ccsdesc[300]{Software and its engineering~Software reliability}
\ccsdesc[300]{Software and its engineering~Software verification}
\ccsdesc[300]{Software and its engineering~Formal software verification}
\ccsdesc[100]{Computing methodologies~Artificial Intelligence}




\maketitle
\section{Introduction}

Informal natural language specifications are omnipresent in modern software. For example, Pfeiffer~\cite{pfeiffer2020constitutes} found natural language documentation in 98\% of over 20,000 GitHub repositories, with 10\% of repository artifacts specifically for documentation. He~\cite{he2019understanding} found over 20\% of non-blank program lines contained in-file comments in their study of 150 of the most starred projects on GitHub.
At the same time, it is well known that software bugs (unexpected exceptions, incorrect output) often arise from the weak association between the intended behavior (documented in natural language) and the behavior of the implementation~\cite{tan2007icomment, tan2012tcomment}. This issue is exacerbated with AI-assisted programming where users generate code from natural language intent~\cite{copilot,codewhisperer,tabnine}, without a good way to ensure their association. Reliably translating informal natural language descriptions to formal specifications could help catch bugs before production and improve trust in AI-generated code~\cite{lahiri2022interactive}.

Current approaches to translating natural language to formal specifications are heuristic-based and either rely on the input being in a structured format~\cite{blasi2018translating, tan2012tcomment} or can only generate a restricted class of specifications (e.g., regarding nullness or exceptions)~\cite{goffi2016automatic, tan2007icomment}. 
Further, most of these approaches are customized for only one specific programming language (such as Java). 
In the past, large-scale neural modeling for the problem of generating specifications has been difficult given the absence of large code corpora with matching natural language intent and corresponding specifications.

Large Language Models (LLMs) have generated a lot of interest in the area of programming owing to their ability to synthesize high-quality code from natural language intent in a surrounding context~\cite{chen2021evaluating, nijkamp2022codegen, li2023starcoder}. 
Given the limitations of the current approaches for translating natural language to formal specifications, we explore the use of LLMs for this problem. 
Even though these LLMs have not seen structured data matching natural language intent to specifications, larger models such as \GPTfour{} have demonstrated ``emergent abilities'' to do well on tasks that they were not explicitly trained for, or unlikely to be common in the training corpus~\cite{wei2022emergent}.
In particular, models such as \GPTfour{} demonstrate capabilities to follow natural language instructions to perform reasoning tasks, for example, through  prompting strategies like few-shot learning~\cite{lampinen2022language}, chain-of-thought~\cite{wei2023chainofthought} and multi-step reasoning~\cite{zhou2023leasttomost}.

In this paper, we explore the feasibility of leveraging LLMs to act as a usable and practical bridge between informal natural language and useful method postconditions. A {\it postcondition} for a method is an assertion that relates the input and output states of the method, and holds true after any successful execution of the method. We assess this ability in a programming language-agnostic way, by targeting postconditions that can be expressed as assertions in the underlying programming language. We term this approach as \NLSpec{} --- leveraging LLMs for the purpose of translating natural language method-level comments to corresponding postconditions. 

\subsection{Motivating Examples}
\label{sec:motivation}

\subsubsection{Formalizing User Intent}

\begin{figure}
    \begin{subfigure}{0.3\textwidth}
        \caption{Programmer intent for a function that removes from a list all instances of numbers that have duplicates.}
        \label{fig:intent}
        \includegraphics[width=\linewidth]{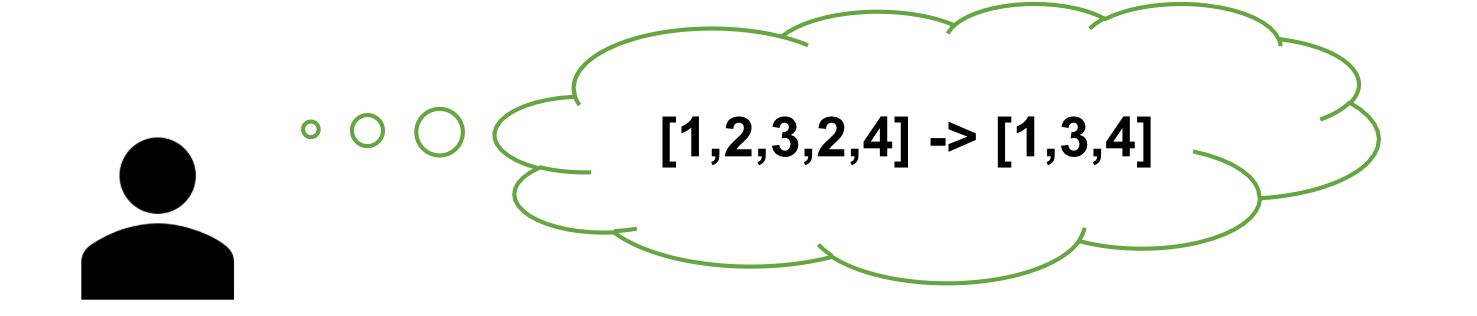}

    \end{subfigure}
    \hspace{2mm}
    \begin{subfigure}{0.67\textwidth}
        \small
        
        \caption{\textit{Ambiguous} natural language specification: it does not specify if all copies or all but one copy of a duplicated element should be removed. In this case, the programmer intends the former. \label{fig:ambiguous}}
        \begin{Python}
  def remove_duplicates(numbers: List[int]):
    """ From a list of integers, remove all elements that occur more than once. Keep order of elements left the same as in the input """
        \end{Python}

    \end{subfigure}
    
    \begin{subfigure}{0.75\linewidth}
         \captionsetup{width=1.32\linewidth}
        \caption{\label{fig:postconditions} Postconditions generated by \GPTfour{}. Note that while both could be correct with a literal reading of the natural language specification, only the second one is correct with respect to developer intent}
            \begin{tabular}{l|c}
                \begin{Python}
  assert len(set(numbers)) == len(set(return_list))
        \end{Python}  &  \color{deepred}\xmark \\
                \begin{Python}
  assert all(numbers.count(i) == 1 for i in return_list)
        \end{Python}           &  \color{deepgreen}\checkmark
            \end{tabular}
      
    \end{subfigure}
    \caption{ 
    \label{fig:motivation} 
    Example of how postconditions could be used to clarify ambiguous natural language specifications.
    }
\end{figure}

Consider the example in Fig.~\ref{fig:motivation}, taken from the popular Python code generation benchmark, 
\humaneval{}~\cite{chen2021evaluating}. A programmer intends a function that removes all numbers with duplicates from a list. For example, given the list \pyinline{[1,2,3,2,4]} the function should return \pyinline{[1,3,4]} without the element \ic{2} because it appears more than once (Fig.~\ref{fig:intent}). The programmer describes the function specification in a docstring (see Fig.~\ref{fig:ambiguous}). However, the natural language specification is ambiguous; 
it does not indicate if all copies of the duplicated elements should be removed, or if one copy should be retained. In this case, the programmer intends the former, however, it is not uncommon to expect that the program should fulfill the latter. 

\begin{figure}
    \begin{subfigure}{0.8\textwidth}
        \caption{Original function plus java doc.}
\begin{Java}
  /** @return a new line with reversed direction.*/
  public Line revert() {
    final Line reverted = new Line(zero, zero.subtract(direction));
    return reverted; }
\end{Java} \label{fig:enter-label}
    \vspace{-9pt}
    \end{subfigure}

    \begin{subfigure}{0.90\textwidth}
        \vspace{+2pt}
        \caption{Bug report (not seen by the LLM) \label{fig:motivationbugreport}}
        \setlength{\FrameSep}{5pt}
        \begin{framed}            \scriptsize
            \color{deepred}\texttt{Line.revert() only maintains $\sim$10 digits for the direction. This becomes an issue when the line's position is evaluated far from the origin.} 
        \end{framed}
    \end{subfigure}

    \begin{subfigure}{0.9\textwidth}
        \caption{\label{fig:bugspec} Two bug-catching post conditions generated by GPT 4}
        \begin{Java}
  // Correct bug-finding postcondition 1
  assert this.direction.negate().equals(returnValue.direction);
  //Correct bug-finding postcondition 2
  assert this.direction.dotProduct(returnVal.direction) == -1 * this.direction.getNormSq();
        \end{Java}
    \end{subfigure}
    
    \caption{ 
    \label{fig:mathDfjExample} Example of how postconditions or other formal specifications of program behavior could catch bugs. This example is a historical bug from \dfj{} (Math-9): the \texttt{Line} constructor does not return a new line with enough precision. The postconditions were generated by \GPTfour{} in our evaluation, and both catch the bug.  
     }
\end{figure}

Figure~\ref{fig:postconditions} contains two postconditions, each satisfying one of the two possible intents of the ambiguous docstring. The programmer writing the \ic{remove\_duplicates} function can verify that the second postcondition ``\pyinline{assert all(numbers.count(i)==1 for i in return\_list})'' correctly matches their intent, by ensuring that all numbers in the returned list occur exactly once in the input list. The first postcondition, however, incorrectly asserts that \emph{\ic{return\_list}} is a set of the input list \emph{\ic{numbers}}. In this example from \humaneval{}, it is not immediately clear at first glance what the user intent is. Generating such postconditions from natural language allows for checkable and unambiguous statements about a program's intended behavior, formalizing a user's intent about a program.

 \subsubsection{Detecting Real-World Functional Bugs}

In practice, postconditions generated in the target programming language can be used in assertions, as demonstrated in fig.~\ref{fig:postconditions}, to check the correctness of a function, enabling the early detection of bugs or violations of a programmer's intent. 

The example in Figure~\ref{fig:mathDfjExample} shows how formal specifications can be used to catch bugs in real-world programs. A bug from the \ic{Apache Commons Math} project, the function \ic{revert()} calls a constructor \ic{Line()} that should return a new Line object with a reversed direction. The bug report~\footnote{\url{https://issues.apache.org/jira//browse/MATH-938}} associated with the issue explains that \ic{revert()} does not maintain enough precision, and fails in certain scenarios. Both of the provided postconditions in Figure~\ref{fig:bugspec} catch the bug by leveraging project-specific context and general mathematical knowledge about the specifications of a reversed line.

\subsection{Overview}
In these examples, we demonstrated that \GPTfour{} can generate postconditions from natural language that closely capture informal intent and also detect program bugs. However, it is unclear to what extent LLMs are capable of the \NLSpec{} problem in general. 
We pose the high-level question: 
\begin{quote}
\textit{Given a natural language description of a method and a candidate postcondition, how do we judge the quality of the postcondition?}
\end{quote}

We attempt to study this question through two high-level research questions:
\begin{itemize}
\setlength{\itemindent}{-1em}
    \item RQ1: \rqa
    \item RQ2: \rqb
\end{itemize}

To answer these questions, we define automated metrics for measuring the usefulness of LLM-generated postconditions, describe different ways to encode the problem statement for an LLM, explore different LLMs, and perform an empirical investigation (both quantitative and qualitative) on benchmarks across multiple programming languages. 
We first define automated evaluation metrics for the correctness and completeness (i.e., the discriminative power) of a postcondition (Section~\ref{subsec:metrics}), and we propose a generic ``prompt'' and variants to transform \NLSpec{} into an input for LLM (Section~\ref{subsec:prompts}). We evaluate RQ1 using a Python programming dataset and present a detailed analysis of generated postconditions quality across different LLMs and prompt variants (Section~\ref{sec:rq1}).
Next, we evaluate RQ2 on a benchmark of real-world Java defects and report on the ability of postconditions to find bugs by distinguishing the fixed version from the buggy version (Section~\ref{sec:rq2}).
Finally, we articulate the limitations (Section~\ref{sec:threats}) and discuss related works (Section~\ref{sec:related}).

\subsection{Contributions}
\begin{itemize}[leftmargin=0.2in]

    \item Evaluating the feasibility of LLMs to facilitate \NLSpec{} via an empirical evaluation of the quality and usefulness of LLM generated postconditions on multiple benchmarks in multiple mainstream programming languages.
    \item A set of metrics (both correctness and completeness) for evaluating natural language generated postconditions, validated through an empirical and qualitative investigation. 
    In particular, we believe this paper is the first to propose the use of LLMs to derive a natural distribution of {\it code mutants} to evaluate the completeness of specifications. 
    \item The finding that with sufficiently robust natural language descriptions, LLMs can use \NLSpec{} to generate \textit{correct}  postconditions with high {\it discriminative power}. We illustrate that with \GPTfour{} we can generate correct postconditions for up to 96\% of problems for our benchmark, \textit{EvalPlus}, with correct postconditions able to discriminate up to 81\% of buggy programs on average.
    \item The finding that LLM-generated \NLSpec{} postconditions are precise enough to capture real-world bugs in large industrial projects; \NLSpec{} postconditions detect 64 historical bugs from 70 buggy methods in industrial-scale Java projects. 
    \item We contribute and release the artifacts of our study including LLM-generated code mutants and postconditions that can be useful for future research in this area.

\end{itemize}
\section{\NLSpec{}: Overall Approach}
\subsection{Problem formulation and metrics}
\label{subsec:metrics}

\newcommand{\nl}{{\it nl}}
\newcommand{\refr}{{\it r}}
\newcommand{\tests}{{\it T}}
\newcommand{\ret}{{\it ret}}
\newcommand{\post}{{\it post}}
\newcommand{\postset}{{\it P}}
\newcommand{\expreval}[2]{\it eval(#1, #2)}
\newcommand{\expr}{\it expr}
\newcommand{\correct}[1]{\it correct_{#1}}
\newcommand{\btrue}{\it true}
\newcommand{\bfalse}{\it false}

We first formalize the \NLSpec{} problem through metrics to evaluate the quality of generated postconditions. 
Consider an example $\langle \nl, \refr, \tests\rangle$, where $\nl$ is the natural language description of a problem, $\refr$ is a reference code implementation, and $\tests$ is a set of test inputs.
For this section, we assume that each test $i \in \tests$ is an input that assigns a value to the input parameters and globals of $\refr$.
We further assume that the reference solution is deterministic and returns a single output value $\ret$ containing the output. 
In this simple setting, it suffices to only have the set of inputs in $\tests$, as the desired output for each input $i$ can be obtained by executing $\refr(i)$.
For the purpose of postcondition generation through an LLM, the set of tests $\tests$ is {\it hidden} from the LLM that generates a postcondition.
The reference implementation $\refr$ may or may not be present during the postcondition generation. 
However, both $\refr$ and $\tests$ are used to define the metrics for the offline evaluation for a benchmark set.

\subsubsection{Test-set correctness} 
\label{sec:correctness}
Given an example $e \doteq \langle \nl, \refr, \tests\rangle$, a candidate postcondition $\post$ is an assertion over the input and output states of $\refr$. 
For an expression $\expr$ and a state $s$ that assigns valuation to variables, let $\expreval{\expr}{s}$ be the result of evaluating $\expr$ after replacing the variables in $\expr$ with their values from $s$. 
A postcondition is {\it correct} if the reference implementation $\refr$ satisfies it for every possible (legal) input.
Therefore, a candidate postcondition $\post$ is correct if for every input $i$, if $\refr(i)$ is the output value, then $\expreval{\post}{(i, \refr(i))}$ is true, where $(i, \refr(i))$ is the joint state of the input parameters and output return variable.
However, such a notion of correctness is difficult to establish in the absence of formal verification tools, and may further require manual effort to establish such proof even for verification-aware languages~\cite{leino2010dafny, swamy2011fstar}.
We take a pragmatic approach, assuming that the test cases in $\tests$ are sufficiently comprehensive to approximate the space of all legal inputs.
Therefore, an expression $\post$ is {\it test-set-correct} w.r.t. $\tests{}$ (denoted as $\correct{\tests}$)  iff $\forall i \in \tests: \expreval{\post}{(i, \refr(i)} == \btrue$.
Henceforth, we may refer to "test-set-correct" as simply correct, since correctness in the remainder of the paper is with respect to the provided tests. 

\newcommand{\passat}[1]{\it pass@{#1}}
\newcommand{\acceptat}[1]{\it accept@{#1}}

Given a set of $m$ postconditions from an LLM, we define a metric $\acceptat{k}$ for $1 \leq k \leq m$ to capture the statistical expected value of containing at least one test-set-correct postcondition while sampling subsets of size $k$ from the set of $m$ conditions.
This is inspired by the $\passat{k}$ metric proposed for evaluating the quality of correctness of generated code given a set of tests~\cite{chen2021evaluating}. 

\newcommand{\codeMutants}{\it CM}

\subsubsection{Test-set completeness for code mutants: \textit{bug-completeness-score}}
\label{sec:mutants}
(Test-set) correctness is a {\it necessary} condition for a valid and useful postcondition, however, it is not {\it sufficient}.
For example, the expression $\btrue$ vacuously satisfies any implementation $r$ for any input $i \in \tests$, and is therefore correct.
The value of a postcondition comes from how well it captures the desired intent expressed in the natural language intent $\nl$.
However, given that $\nl$ is informal, we cannot establish a check to ensure the association. 
Instead, we leverage the reference implementation and tests as the source of {\it ground truth} for what the user intends.
However, this again poses the problem that the most desired postcondition is simply the {\it strongest postcondition} of $\refr$ program, which is computationally intractable~\cite{dijkstra1990strongest}. 
Instead, we use a concept of {\it completeness} that measures the degree to which the postcondition distinguishes the reference implementation $\refr$ from other incorrect implementations.

Inspired by mutation-testing literature (c.f. \citet{jia2010analysis}) that assigns a  score to a test $t$ based on the fraction of code mutants ``killed'' or distinguished under $t$, we assign a measure of {\it bug-completeness} to a postcondition $\post$ as the fraction of code mutants that can be distinguished given the set of tests $\tests$.
Unlike traditional mutation testing, we parameterize completeness with a  {\it semantically distinct} code mutant set $\codeMutants$ that are guaranteed to differ from $\refr$ (and from each other) on at least one test in $\tests$.
In other words, for each $c \in \codeMutants$, there exists a test input $i \in \tests$ that distinguishes from $\refr$ (i.e., $\refr(i) \neq c(i)$) and (a possibly different) $i$ that distinguishes from any other $c' \in \codeMutants \setminus \{c\}$ (i.e., $c(i) \neq c'(i))$. 
Given an example $e \doteq \langle \nl, \refr, \tests\rangle$, a $\correct{\tests}$ postcondition $\post$ and a set of distinct code mutants $\codeMutants$, we define the \textit{bug-completeness-score} 
of $\post$ as:
$$
\textit{bug-completeness-score}(\post, \codeMutants, \tests) \doteq |\{c \in \codeMutants \ | \ \exists i \in \tests: \expreval{post}{(i, c(i))} == \bfalse\}|/|\codeMutants|
$$
In other words, \textit{bug-completeness-score} measures the fraction of code mutants that fail the correct postcondition. If the \textit{bug-completeness-score} of a postcondition is 1, we say that the postcondition is \textit{bug-complete}.
One can easily lift the idea to the completeness of a set postconditions $\postset$ by taking a union of all the code mutants ``killed'' using all the correct postconditions in the set:
$$
\textit{bug-completeness-score}(\postset,\codeMutants, \tests) \doteq |\bigcup_{\post \in \postset} \{c \in \codeMutants \ | \ \exists i \in \tests: \expreval{post}{(i, c(i))} == \bfalse\}|/|\codeMutants|
$$
\newcommand{\impls}{\it Impls}

We now discuss why we use a parameterized set of code mutants instead of creating variants of $\refr$ by mutating different operators. 
We believe that such a fixed set of mutation operators does not approximate real-world bugs for two reasons: (a) first, since code mutants only differ from the reference implementation in one or two operators at a time, it may not cover mutations that are further away in the edit distance, and (b) it may not cover subtle bugs that a human would write using different syntactic constructs (e.g., a while loop instead of a for loop) or APIs. 
We propose the use of LLMs to sample mutants from the natural distribution of implementations to the problem described by the natural language intent $\nl$. 
In other words, we enumerate a set of likely implementations $\impls$ for $\nl$ using a LLM (such as GPT-4), and define $\codeMutants$ to be the subset of $\impls$ that differ from $\refr$ on at least one test $i \in \tests$, and also pairwise distinct in terms of the tests in $\tests$. 

\newcommand{\testMutants}{\it TestMutants}

\subsection{Prompt Design for LLM-based Postcondition Generation}
\label{subsec:prompts}
\begin{figure}
    \centering
    \includegraphics[width=0.75\linewidth]{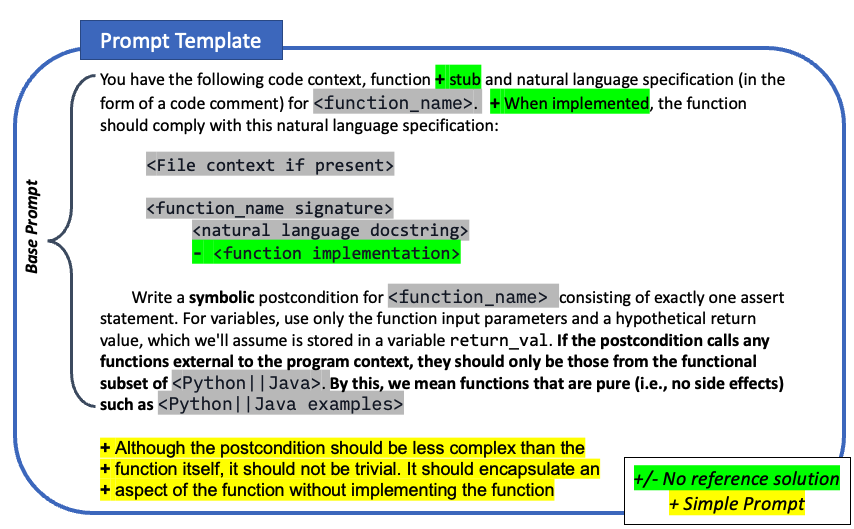}
    \caption{Prompt template for generating postconditions from natural language via chat models (including changes needed for the \ic{simple} and \textit{no reference} variations). We found that the \textbf{bold} text greatly improved the quality of the postconditions: without it, the model tended to return point-based tests or code blocks with side effects. While modified here slightly for clarity, our full prompts are included in our replication package.}
     \label{fig:prompts}
\end{figure}

LLM performance has been shown to be impacted by small changes in prompts for the same problem task, and designing the optimal prompt is not always a straightforward task.  We explore several prompt templates, i.e. varied textual representations of the problem description $\nl,$ and reference solution $ \refr$, optimizing for a number of outcomes. First, the prompts should work with \textit{chat-based models}, and the generated postconditions should be \textit{symbolic} (e.g., not point-wise tests), directly executable, and side-effect free.  Also, the prompt should encourage the LLM to produce expressions that are syntactically and semantically valid while being as programming language agnostic as possible. 
Several prompt iterations were considered until we observed satisfactory performance on a subset of example problems, though we acknowledge further prompt tuning may result in different outcomes. Figure~\ref{fig:prompts} outlines our prompt template. This template shows four possible prompt iterations along two orthogonal axes (a) whether the reference $\refr$ is included, and (b) requested postcondition complexity. We now discuss each axis in more detail.

\textit{Including reference code:} Our default prompt includes only the $\nl$, not the reference code $\refr$. 
This is useful for {\it specification-driven AI-based programming} scenarios~\cite{lahiri2022interactive} where the user first accepts a few specifications that are used to constrain AI generated code suggestions. 
However, we also provide a prompt variant that includes the reference code $\refr$ along with $\nl$. 
This allows us to assess if natural language alone can be as effective as code in conveying programming intent to an LLM. 

\textit{Postcondition complexity:} we also consider a \ic{simple} variation of the prompt that explicitly instructs the LLM to generate postconditions that capture an aspect of a function, rather than the whole function. 
We include this variation as we noticed that when using the \ic{base} prompt, LLMs have a tendency to construct complex postconditions, often striving for a fully functional implementation of the problem. 
While useful, we observe these complex postconditions are more likely to be incorrect. To understand our motivation for including both prompt variations, Fig.~\ref{fig:simplevsbase} compares postconditions produced by the \ic{base} and \ic{simple} prompt for a problem from the \humaneval{} benchmark.

\begin{figure}
    \begin{subfigure}{0.95\textwidth}

        \caption{Informal natural language specification for problem 12 from \humaneval{}}
        \setlength{\FrameSep}{5pt}
        \begin{framed}            \scriptsize
\color{deepred}
\texttt{Given a string text, replace all spaces in it with underscores, and if a string has more than 2 consecutive spaces, then replace all consecutive spaces with - . For example: fix\_spaces(`` Example 1'') == ``\_Example\_1'', fix\_spaces(`` Example   2'') == ``\_Example-2''
    } \end{framed}
    \end{subfigure}
    \begin{subfigure}{0.95\textwidth}

        \caption{\ic{base} vs. \ic{simple}: the \ic{base} postcondition tries to capture all intended functionality, but does so incorrectly. The \ic{simple} postcondition is less complex (capturing less functionality), but is correct.}
\begin{Python}
  # Base prompt: postcondition that incorrectly attempts to fully specify the problem
  assert all(map(lambda x: x == "_" or x == "-", re.split(r'\w+', return_value))) and "  " not in return_value and "__" not in return_value and "--" not in return_value
  
  # Simple prompt: postcondition checks that return_value does not contain any  spaces.
  assert not re.search(r' {1,}', return_value), "The return value contains one or more spaces"
\end{Python}
        \label{fig:enter-label}
    \end{subfigure}
    
    \caption{ 
    \label{fig:simplevsbase} Example of how the \ic{base} and \ic{simple} prompt variations can impact postcondition construction. Both postconditions were generated for \humaneval{} problem 12 using \GPTfour{}.
     }
\end{figure}

We combine these two prompt variations into four distinct prompts in our evaluation:

\begin{enumerate}
    \item \ic{Base} prompt with only natural language description $\nl$ (no reference solution $\refr$)
    \item \ic{Base} prompt with both reference $\refr$ and natural language description $\nl$
    \item \ic{Simple} prompt with only natural language description $\nl$ (no reference solution $\refr$)
    \item \ic{Simple} prompt with both the reference $\refr$  and natural language  description $\nl$
\end{enumerate}

\section{RQ1: \rqa}
\label{sec:rq1}

To assess if LLMs can generate high-quality postconditions that capture and formalize intent, we report a detailed empirical study of LLM-generated postconditions on a popular benchmark.

\subsection{RQ1 Experimental Setup}

\subsubsection{Evaluation Benchmark}

\label{subsec:humanEvalBenchmark}

We use the benchmark \humanevalplus{}, which includes 164 Python  problems, each with an associated function stub and natural language description in the form of a Python docstring, a reference implementation, and validation tests~\cite{liu2023IsYourCode}. 
\humanevalplus{} is an update to the popular \humaneval{} benchmark~\cite{chen2021evaluating}, containing the same problems but with the addition of a more extensive test suite (775 tests per problem on average). 
We choose \humanevalplus{} because each example has (a)  a descriptive natural language intent, (b) a set of extensive test inputs, and (c) a unique reference solution. Using these three components, we can evaluate if a postcondition formalizes the user intent expressed in the natural langauge docstring, $\nl$, while also satisfying the reference solution.

\subsubsection{Large Language Models}

We generate postconditions using three recent chat-based models, including both closed and open-source approaches, that have demonstrated state-of-the-art performance on various programming tasks:
\begin{itemize}[leftmargin=0.2in]
    \item \emph{OpenAI: \textbf{\GPTthree{}} and \textbf{\GPTfour{}}} are based on the pre-trained GPT-3 model, which is further fine-tuned using Reinforcement Learning with Human Feedback (RLHF)~\cite{ouyang2022training}. While \GPTthree{} and \GPTfour{} are not explicitly fine-tuned for code generation, they have demonstrated strong capabilities on several related tasks~\cite{olausson2023demystifying, fakhoury2023towards}. We use OpenAI APIs for the \ic{gpt-3.5-turbo} and \ic{gpt-4} endpoints.

    \item \emph{The BigCode Project: \textbf{\opensource{}.}} StarCoder \cite{li2023starcoder} is an open-access 16B parameter model pre-trained on The Stack~\cite{kocetkov2022stack}, one trillion tokens sourced from 80+ programming languages, GitHub issues, Git commits, and Jupyter notebook. We use \emph{StarChat}~\footnote{\href{https://huggingface.co/HuggingFaceH4/starchat-alpha}{HuggingFace model identifier {\tt HuggingFaceH4/starchat-alpha}}}, a version of StarCoder fine-tuned for assisting coding. 
    StarChat is currently one of the few open-access chat model alternatives to \GPTthree{} and \GPTfour{}, permitting replication of and comparison with our results. This model allows us to use the same prompt we used for the OpenAI models, rendering a fairer comparison. 
\end{itemize}

\subsubsection{Postcondition Generation}

\label{subsec:MethodsPostconditionGenderation}

For each \humanevalplus{} problem, we generate 10 postconditions for each of the 4 prompt variants (Section~\ref{subsec:prompts}) per LLM model. 
We use a temperature of 0.7 as it is the default for both \GPTthree{} and \GPTfour{}, and has been found to be a reasonable temperature for code generation tasks.\footnote{We also considered additional temperatures of 0.2 and 1.2, and we include the results in our replication package. However, since high-level trends were the same regardless of temperature, we only report results of 0.7 here for clarity.} 
As we consider four prompt variants, we generate 40 postconditions per problem per model. This results in 19,680 postconditions across all variants, models, and \humanevalplus{} problems.

\subsubsection{Code Mutant Generation}
\label{sec:codeMutants}
To generate the set of code mutants $\codeMutants$ needed for \textit{bug-completeness}, we use an LLM (\GPTthree{} with temperature 0.9) to generate a set of codes that satisfy the natural language intent $\nl$, and filter the ones that fail at least one test in $\tests$.
We generate 200 code solutions to each problem and then save only those that fail the test suite. 
We term these bugs as \textit{natural} code mutants, as they represent natural yet buggy implementations for the problem description. 
However, we noticed that for some examples, the number of such natural code mutants is fairly small. 
To amplify the set of buggy codes, we generate 200 additional buggy codes by explicitly instructing \GPTthree{} to include an error in its solution. 
As mentioned in Section~\ref{sec:mutants}, we only retain distinct buggy codes so that no two mutants fail the same set of tests. 
The number of unique buggy codes varies per problem, ranging from 4 to 233 with a median of 55. While we combine the two bug sources, we also consider the \textit{natural mutants} alone in our evaluation to see if the source of the bug impacts the efficacy of our metrics. We make available the set of mutants for use by the broader research community and to support the reproducibility of our results.

\subsection{RQ1-Results: Do LLM-generated postconditions formalize user intent?}

We now discuss the results of our empirical and qualitative evaluations, structured around postcondition correctness, postcondition completeness, and qualitative insights.

\subsubsection{Postcondition Correctness}
\label{subsec:humanEvalCorrectness}

\begin{table}[]
\scriptsize
\begin{tabular}{@{}lrcrrrr@{}}
\toprule 
\\
Model      & Prompt& \multicolumn{1}{l}{\multirow{-2.5}{*}{\begin{tabular}[c]{@{}c@{}}Prompt has:\\ NL Only=\color{deepred}\xmark\\ ref code=\color{deepgreen}\checkmark\end{tabular}}} & \begin{tabular}[c]{@{}r@{}}Accept\\  @ 1\end{tabular} & \begin{tabular}[c]{@{}r@{}}Accept \\ @ 5\end{tabular} & \begin{tabular}[c]{@{}r@{}}Accept \\ @ 10\end{tabular} & \begin{tabular}[c]{@{}r@{}}x/164\\ correct\end{tabular} \\ \midrule

\GPTthree{} &\ic{base} &\color{deepred}\xmark       & \cellcolor[HTML]{FFFFFF}0.46  & \cellcolor[HTML]{FFFFFF}0.80  & \cellcolor[HTML]{FFFFFF}0.87  & \cellcolor[HTML]{FFFFFF}143 \\

\GPTthree{} &\ic{base} &\color{deepgreen}\checkmark & \cellcolor[HTML]{FCFBFD}0.49  & \cellcolor[HTML]{FDFCFE}0.81  & \cellcolor[HTML]{FBFAFD}\textbf{0.88} & \cellcolor[HTML]{FAF9FD}\textbf{145} \\

\GPTthree{} & \ic{simple} & \color{deepred}\xmark & \cellcolor[HTML]{F4F2F9}0.55 & \cellcolor[HTML]{FAF9FC}\textbf{0.82} & \cellcolor[HTML]{FFFFFF}0.87 & \cellcolor[HTML]{FFFFFF}143 \\

\GPTthree{} & \ic{simple} & \color{deepgreen}\checkmark & \cellcolor[HTML]{F3F1F8}\textbf{0.56} & \cellcolor[HTML]{FAF9FC}\textbf{0.82} & \cellcolor[HTML]{FBFAFD}\textbf{0.88} & \cellcolor[HTML]{FDFCFE}144 \\

\midrule
\GPTfour{} & \ic{base}   & \color{deepred}\xmark & \cellcolor[HTML]{EBE7F3}0.63  & \cellcolor[HTML]{F7F6FB}0.83   & \cellcolor[HTML]{FBFAFD}0.88 & \cellcolor[HTML]{FDFCFE}144  \\

\GPTfour{} & \ic{base}   & \color{deepgreen}\checkmark                             & \cellcolor[HTML]{E1DBEE}0.71                          & \cellcolor[HTML]{E7E3F1}0.89                      & \cellcolor[HTML]{EFEBF6}0.91                           & \cellcolor[HTML]{EEEAF5}150                             \\

\GPTfour{} & \ic{simple} & \color{deepred}\xmark & \cellcolor[HTML]{D9D2E9}\textbf{0.77}                 & \cellcolor[HTML]{D9D2E9}\textbf{0.94}                 & \cellcolor[HTML]{D9D2E9}\textbf{0.96}                  & \cellcolor[HTML]{D9D2E9}\textbf{158}                    \\

\GPTfour{} & \ic{simple} & \color{deepgreen}\checkmark                             & \cellcolor[HTML]{DBD4EA}0.76                          & \cellcolor[HTML]{DFD9ED}0.92                          & \cellcolor[HTML]{D9D2E9}\textbf{0.96}                  & \cellcolor[HTML]{DCD5EB}157                             \\

\midrule

\opensource{} & \ic{base}   & \color{deepred}\xmark &  0.21 &  0.61 &  0.82 & 134\\

\opensource{} & \ic{base}   & \color{deepgreen}\checkmark & 0.20 &  0.59  &  0.77   & 126 \\

\opensource{} & \ic{simple} & \color{deepred}\xmark & \textbf{ 0.25} & \textbf{0.69 }& \cellcolor[HTML]{FBFAFD} 0.85 &  139 \\

\opensource{} & \ic{simple} & \color{deepgreen}\checkmark   &     0.23   &  0.67   &  \cellcolor[HTML]{FBFAFD} \textbf{0.86} &  \textbf{141} \\ 
\bottomrule
\end{tabular}

\caption{\label{tab:humanEvalCorrectness} Test-set correctness on \humanevalplus{} for three models (\GPTthree{}, \GPTfour{}, and \opensource{}), differing prompt complexities (\ic{base} vs. \ic{simple}), and including or omitting the reference solution in the prompt. Darker highlighted cells are more correct. Bolded values are the largest for a specific model.} 
\end{table}%

Table~\ref{tab:humanEvalCorrectness} has our \textit{test-set-correctness} (Section~\ref{sec:correctness}) results. 
Overall, we find that for \humanevalplus{}, LLM-generated postconditions are likely to be test-set correct; in our best-performing prompt variation, 77\% of postconditions were test-set-correct and a test-set-correct postcondition was generated for 96\% of problems (158/164). As we show later in this section, test-set-correctness on \humanevalplus{} largely corresponds to true correctness. Our  results indicate that LLMs have the potential to reliably generate correct postconditions from natural language specifications.

Regardless of the prompt variation, \GPTfour{} postconditions were the most likely to be correct ($0.63 \leq $ \texttt{accept@1} $\leq 0.77$) followed by \GPTthree{} ($0.46 \leq $ \texttt{accept@1} $\leq 0.56$). 
\opensource{} postconditions were consistently the least correct, with \texttt{accept@1} between 0.21 and 0.25. 
While the raw number of correct \opensource{} postconditions was low, the number of benchmark problems with at least one correct postcondition was relatively high, ranging from 78\% to 86\% depending on the prompt. 

As described in section~\ref{subsec:prompts}, we consider both a \ic{base} postcondition prompt and a \ic{simple} prompt for generating simpler postconditions that capture only an aspect of program behavior. 
Regardless of LLM model, \ic{simple} postconditions are more likely to be correct than \ic{base}  postconditions. Using a paired students $t$-test~\cite{hsu2014paired} between \texttt{accept@1} ablation pairs where the only difference is the prompt complexity type, \ic{simple} prompt postconditions are significantly more likely to be correct with $p=0.008$, a large effect (standardized Cohen's $d=1.73$). This indicates that when prioritizing correctness, 
using a prompt that explicitly asks for simpler postconditions improves the result. 

We also compared the efficacy of generating postconditions from natural language alone to generating when a reference solution is included in the prompt. We did not observe a significant difference in \texttt{accept@1} between postconditions generated with natural language specifications alone and those including a reference solution ($p = 0.42$). This indicates that the presence of a reference solution does not necessarily enhance postcondition correctness when compared to natural language alone. Therefore, it might be feasible to rely solely on natural language intent (when comprehensive enough) without needing to provide a reference solution.

\textit{Correctness false positives.} \humanevalplus{} has more comprehensive tests than its predecessor \humaneval{}, but it is still possible that the tests do not capture all possible inputs. If so, our test-set correctness metric may have false positives. 
We validate our metric for \humanevalplus{}: we find only one problem (\# 122) with false positives relating to negative inputs. Overall, only 1.1\% of the 900 postconditions that we manually annotated were effected (see Section~\ref{sec:taxonomy} for our annotation process). In contrast, we also compare our results to the hypothetical results if using \humaneval{} (which has the same problems, but many fewer tests). The \humaneval{} results contain 7\% false positives for \texttt{accept@10} for \GPTfour{}, much higher than our results. Thus, we find that test-set correctness is a reasonable approximation for true correctness on \humanevalplus{} (but perhaps less so for \humaneval{}).

\begin{tcolorbox}[title=RQ1 Summary: Postcondition Correctness on \humanevalplus{}]
    \noindent On \humanevalplus{}, LLMs are good at producing \textit{correct} postconditions from informal natural language specifications. All prompt variants generate a correct postcondition for at least 77\% and up to 96\% of problems. \GPTfour{} consistently outperformed \GPTthree{} and \opensource{}. Explicitly asking for \ic{simple} postconditions leads to more correct postconditions. However, we do not observe a significant difference between including or omitting a reference solution in the prompt; using natural language (when descriptive) alone can be just as powerful.
\end{tcolorbox}

\subsubsection{Postcondition Completeness} 
\label{subsec:humanEvalCompleteness}
While our test-set correctness results are encouraging, test-set correctness is necessary but not sufficient for assessing if a postcondition meaningfully captures the natural language specification. To capture a notion of completeness, we measure \textit{bug-completeness} for all test-set correct postconditions (see Section~\ref{subsec:metrics}). Table~\ref{tab:humanEvalCompleteness} contains our results. We report both the percentage of postconditions that are \textit{bug-complete} (kill all unique code mutants) and the average  \textit{bug-completeness score} (fraction of code mutants killed). 
The results indicate the \GPTfour{} postconditions can kill all the code mutants for up to 
$62.2\%$
of examples in \humanevalplus{}. 
Overall, both \GPTthree{} and \GPTfour{} generate relatively bug-complete postconditions, with average scores of up to 0.76 and 0.85 respectively. That is, the average correct postcondition generated by these models can discriminate over three-quarters of unique buggy code mutants. The bug-completeness scores for \opensource{} were lower but still substantial, catching up to one-third of mutants. Our bug-completeness results suggest that LLMs, especially the advanced models like \GPTthree{} and \GPTfour{}, can use natural language to produce postconditions that meaningfully capture desired aspects of program behavior.

\begin{table}[t!]
\scriptsize
\setlength{\tabcolsep}{3.2pt}.
\begin{tabular}{lrcrrrrr}
\hline
 &  & \multicolumn{1}{l}{\multirow{-1.5}{*}{\begin{tabular}[c]{@{}c@{}}Prompt has:\\ NL Only=\color{deepred}\xmark\\ ref code=\color{deepgreen}\checkmark\end{tabular}}}  &  & \multicolumn{1}{l}{\multirow{-1.5}{*}{\begin{tabular}[c]{@{}r@{}}\% problems\\ with bug-\\complete\end{tabular}}} &  \multicolumn{1}{l}{\multirow{-1.5}{*}{\begin{tabular}[c]{@{}r@{}}\% problems \\ union bug-\\complete\end{tabular}}} & \multicolumn{2}{c}{\begin{tabular}[c]{@{}c@{}}Avg bug-completeness-score \\ for correct postconditions \end{tabular}} \\
\multirow{-2.7}{*}{Model} & \multirow{-2.7}{*}{Prompt} & & \multirow{-2.7}{*}{\begin{tabular}[c]{@{}r@{}}\% bug-\\ complete\end{tabular}} & & & \textit{Natural bugs}& \textit{All bugs}\\ \hline
\GPTthree & \ic{base} & \color{deepred}\xmark & \cellcolor[HTML]{F4F2F9}15.4& \cellcolor[HTML]{EFECF6}42.1& \cellcolor[HTML]{EDEAF5}48.2& \cellcolor[HTML]{E8E4F2}0.62& \cellcolor[HTML]{EBE8F4}0.72\\

\GPTthree & \ic{base} & \color{deepgreen}\checkmark & \cellcolor[HTML]{F0EDF6}\textbf{18.5}& \cellcolor[HTML]{ECE8F4}\textbf{47.0}& \cellcolor[HTML]{EEEBF5}\textbf{49.4}& \cellcolor[HTML]{E3DEEF}\textbf{0.70}& \cellcolor[HTML]{E4DFF0}\textbf{0.76}\\

\GPTthree & \ic{simple} & \color{deepred}\xmark & \cellcolor[HTML]{FFFFFF}8.1& \cellcolor[HTML]{FEFEFF}29.3& \cellcolor[HTML]{FCFCFE}33.5& \cellcolor[HTML]{FDFCFE}0.44& \cellcolor[HTML]{FBFBFD}0.55\\

\GPTthree & \ic{simple}  & \color{deepgreen}\checkmark & \cellcolor[HTML]{F6F5FA}14.0& \cellcolor[HTML]{F4F2F9}37.2& \cellcolor[HTML]{F4F2F9}41.5& \cellcolor[HTML]{F3F0F8}0.58& \cellcolor[HTML]{EFECF6}0.62\\ \hline

\GPTfour & \ic{base} & \color{deepred}\xmark & \cellcolor[HTML]{DAD3EA}\textbf{35.1}& \cellcolor[HTML]{DDD7EC}\textbf{61.6}& \cellcolor[HTML]{E1DBEE}\textbf{62.2}& \cellcolor[HTML]{D9D2E9} \textbf{0.81}& \cellcolor[HTML]{D9D2E9}\textbf{0.85}\\

\GPTfour & \ic{base} & \color{deepgreen}\checkmark & \cellcolor[HTML]{D9D2E9}34.9& \cellcolor[HTML]{E1DBEE}58.0& \cellcolor[HTML]{DFD9ED}61.6& \cellcolor[HTML]{DCD5EB} 0.78& \cellcolor[HTML]{DCD5EB}0.82\\

\GPTfour & \ic{simple} & \color{deepred}\xmark & \cellcolor[HTML]{FEFDFE}9.2& \cellcolor[HTML]{FFFFFF}26.2& \cellcolor[HTML]{FFFFFF}29.3& \cellcolor[HTML]{FFFFFF}0.40& \cellcolor[HTML]{FFFFFF}0.52\\

\GPTfour & \ic{simple}  & \color{deepgreen}\checkmark & \cellcolor[HTML]{FEFEFF}8.9& \cellcolor[HTML]{FEFEFF}29.3& \cellcolor[HTML]{FAF9FC}36.0& \cellcolor[HTML]{FBFAFD}0.47& \cellcolor[HTML]{F9F8FC} 0.56\\ \hline

\opensource & \ic{base} & \color{deepred}\xmark & 0.8& 7.3& 8.5& 0.13&  0.24\\

\opensource & \ic{base} & \color{deepgreen}\checkmark &  1.4& 9.1& 11.0& \textbf{0.23}&   0.30\\

\opensource & \ic{simple}  & \color{deepred}\xmark & 1.5&  6.7&  7.3&  0.16&   0.24\\

\opensource & \ic{simple}  & \color{deepgreen}\checkmark & \textbf{3.0}&  \textbf{17.1}&  \textbf{17.7}&  0.23& \textbf{0.36}\\ \hline
\end{tabular}
\caption{\label{tab:humanEvalCompleteness}Table of bug-completeness for \humanevalplus{}. \texttt{\% bug-complete} is the \% of postconditions that detect all buggy code mutants. \texttt{\% problems with bug-complete} is the \% of all \humanevalplus{} problems with at least one \texttt{bug-complete} postcondition. \texttt{\% problems union bug-complete} is the \% of problems where the union of correct postconditions is \texttt{bug-complete}. Finally, the last two columns are the average \texttt{bug-completeness-score}, a fraction between 0 and 1, for all correct postconditions, normalized by \humanevalplus{} problem. We report this for both \textit{natural} and \textit{all} generated code mutants. Bolded values are the largest value per column per model.}
\end{table} 

In contrast to the correctness results (Section~\ref{subsec:humanEvalCorrectness}), \ic{base} postconditions generally have higher bug-completeness scores than \ic{simple} postconditions (up to a 30\% difference). 
This trend hints that the \ic{simple} prompt may generate more correct postconditions at the expense of bug-catching power.  Even so, as shown in Fig.~\ref{fig:simplevsbase}, \ic{simple} postconditions still meaningfully capture aspects of program behavior: \ic{simple} \GPTfour{} and \GPTthree{} postconditions discriminate over half of unique buggy mutants.

We also compare the bug-completeness of postconditions generated from natural language intent alone to those generated with a reference solution in the prompt. While the difference was not quite significant, the average bug-completeness score was 5\% higher for the case with the reference code included ($p=0.06$). From our qualitative investigation, this seems to be caused by an increase in the number of postconditions that are functional re-implementations of the reference solution. 

\textit{Natural vs. Artificial bugs.} To help validate our proposed bug-completeness metric (see Section~\ref{subsec:metrics}), we examine the impact of using \textit{natural} or \textit{artificial} LLM code generation bugs. As shown in Table~\ref{tab:humanEvalCompleteness}, our completeness metric was consistently (though not always substantially) lower when only considering natural bugs; naturally occurring LLM code generation bugs are harder to kill via \NLSpec{} than artificially seeded bugs. This finding highlights a potential limitation in using artificially seeded faults to assess postcondition correctness as it may artificially inflate the metric. However, generating unique natural bugs is more expensive than using artificial bugs. To ensure metric robustness, augmenting the evaluation metric with artificial bugs may still be useful.
 
\begin{tcolorbox}[title=RQ1 summary: Postcondition Completeness on \humanevalplus{}]
    \noindent We find that for the benchmark \humanevalplus{}, \NLSpec{} postconditions generated by \GPTthree{} and \GPTfour{} can meaningfully capture program intent especially when using our \ic{base} prompt: the average correct postcondition generated by these models can discriminate three-quarters of unique buggy code mutants depending on the prompt variation.
\end{tcolorbox}

\subsubsection{Qualitative Analysis of Generated Postconditions}
\label{sec:taxonomy}
Evaluating postcondition correctness and completeness tells us how well LLMs can generate specifications that capture the programs intent, however it does not give us insight into the kinds of generated specifications, and how they differ in terms of performance. 
We ask two questions:  
1) \textbf{Are there patterns within LLM generated postconditions} and 2) \textbf{How do these categories differ in terms of correctness and completeness?} These insights can help to inform future improvements around LLMs generated specifications, and may guide ranking or selection strategies when using generated postconditions in practice. 

To determine what program aspects \NLSpec{} postconditions verify, we conduct a manual qualitative analysis. We first select 230 postconditions generated for 23 \humanevalplus{} problems. We use the best-performing prompt version for correctness: \GPTfour{} with the \ic{simple} prompt and no reference solution. The first two authors developed a set of qualitative coding categories for postcondition structure and jointly labeled all 230 postconditions. The first author then used this  set of categories to label an additional 670 postconditions for a total of 900 labeled postconditions from 139 \humanevalplus{} problems 
We present these categories in Table~\ref{tab:taxonomy} and report prevalence and completeness.

\begin{table}[]
\footnotesize
\caption{\label{tab:taxonomy} Atomic categories of \NLSpec{} postconditions  \ifbool{longVersion}{\textit{Composable Checkers} and are postcondition building blocks, often conjoined by the LLM via \textit{logical and}. \textit{Complete Specification Types} are high-level patterns of complete postconditions.}{that are often combined by LLMs via \texttt{\&\&} (logical and).} \texttt{return\_val} refers to the function's return value. 
\textit{\% test-set correct} and \textit{bug-completeness} columns are defined in Section~\ref{subsec:metrics}. Example postconditions are adapted from our \humanevalplus{} results, only modified for space. }
\setlength{\tabcolsep}{1.5pt}
\scriptsize{
\begin{tabular}{l l c r}
\toprule
\\ \\
Category \hspace{56pt} & Example Postconditon                        & \% Prevalent \hspace{1pt} & \multirow{-3}{*}{\begin{tabular}[r]{@{}r@{}} Avg. Bug-\\ complete-score\\ (\textit{Natural}/\textit{All})\end{tabular}} \\ 
        \midrule
      
         \ifbool{longVersion}{                
\textit{Composable Checkers} \\
\midrule }{}
Type Check                      & \pyinline{isinstance(return\_val, int)}  & 47.4 & \textit{0.14 / 0.27} \\
Format Check             & \pyinline{return\_val.startswith("ab")}  & 11.2  & 0.43 / 0.57  \\
Arithmetic Bounds          & \pyinline{return\_val >= 0} &     30.8   & 0.23 / 0.34  \\
Arithmetic Equality        & \pyinline{return\_val[0] == 2 * input_val} & 17.5   &\textbf{ 0.82 / 0.89}  \\
Container Property                &\pyinline{len(return_val) > len(input_val)}                                                  &      27.0                  &         0.45 / 0.57                         \\
Element Property        &  \pyinline{return_val[0] \% 2 == 0 }   & 12.6 & 0.39 / 0.53\\
Forall-Element Property  & \pyinline{all(ch.isalpha() for ch in return_val)} & 8.3 & 0.23 / 0.44\\ 
Implication        &\pyinline{(return_val==False) if 'A' not in string} & 12.7 & 0.58 / 0.64 \\
Null Check        &\pyinline{return_val is not None}   &   4.4 &    0.40 / 0.50 \\ \midrule
Average & &  &  0.32 / 0.46\\
\ifbool{longVersion}{
\midrule
\textit{Complete Specification Types}\\
\midrule
\multirow{2}{*}{\begin{tabular}[l]{@{}l@{}}Full Checker \end{tabular}}        &\multicolumn{1}{r}{\multirow{2}{*}{\begin{tabular}[r]{@{}r@{}} \pyinline{assert len(r_val) == n and r_val[0] == n /} \\ \pyinline{ and all(r_val[i] < r_val[i -1] for ...}\end{tabular}}}       &  --   &    100.0   \\
\\
Reverse Checker  &\pyinline{assert ((8*r_val + 1)**0.5 - 1) / 2 == in_val}       & --    &    100.0   \\
Functional Implementation  &\pyinline{assert r_val == sorted(list(set(in_val))}  &  88.1  & 100.0  \\
}{}
\bottomrule
\end{tabular}
}
\end{table}
From the classification process, we observe that postconditions can take the form of either atomic or conjoined statements. For example, an LLM may generate a single postcondition that checks several distinct properties about a program, conjoined with logical \texttt{\&\&} operators. Results of the classification process show that 33\% of LLM-generated postconditions  consists of multiple {\it atomic postconditions}, conjoined using \texttt{\&\&} (logical and). 

We categorize nine basic types of atomic properties. Table~\ref{tab:taxonomy} contains an example of each, along with its dataset prevalence and completeness measures. Prevalence is counted across both atomic and conjoined statements, e.g. if an assertion conjoins specifications across two categories, both are counted. As a result, prevalence adds to over 100\%.
\texttt{Type Checks} enforce a constraint on the type of a return value using \texttt{isinstanace}. 
\texttt{Format Checks} ensure that the return value follows a certain string format constraint. 
\texttt{Arithmetic Bounds} and \texttt{Arithmetic Equality} enforce a numeric bounding or equality constraint against another expression. 
\texttt{Container Property} checks an aspect of a complex type or object (e.g., the length of an array).  
\texttt{Element Property} and \texttt{Forall-Element Property} enforce some  constraint on one or all elements of a collection. 
\texttt{Implications} include conditional logic, and \texttt{Null Check} ensures that the return value is not \texttt{None}. 

We did not observe a significant relation between postcondition type and correctness. However, we do observe significant differences in bug-completeness across categories. For example, postconditions labeled as \texttt{Type Checks}, i.e. specifications enforcing the type of the return value, were the weakest, only killing 27\% of  bugs on average. This difference was particularly pronounced for \textit{natural bugs} (see Sections~\ref{subsec:metrics} and~\ref{sec:codeMutants}), where \texttt{Type Checkers} only killed 14\% of bugs on average. Interestingly, \texttt{Type Checks} are also the most prevalent category, indicating LLM preference towards generating such constraints. {Low completeness scores indicate that, for the studied dataset, type-mismatch errors is not a common bug source. This may be explained by the inclusion of type hints in the \humanevalplus{} dataset, which appear in function stubs provided to the LLM.} 

On the other hand, \texttt{Arithmetic Equality} checks, i.e. specifications that assert that parts of the return value must be equivalent to another expression, provide a strong postcondition. %
On average, this category of postcondition kills 89\% of all bugs and appears in 17.5\% of labeled postconditions. 

Using our categorization, we can partially explain the lower completeness scores of \opensource{} postconditions in section~\ref{subsec:humanEvalCompleteness}. 
While we do not perform a systematic qualitative analysis, we observe that the majority of correct \opensource{} postconditions are atomic \texttt{Type Checks}, which is the weakest postcondition type (see Table~\ref{tab:taxonomy}). This hypothesis is also validated by results of \GPTfour{}, where in contrast, only 16\% of generated postconditions are atomic \texttt{Type Checks} alone. Instead, the majority of \texttt{Type Checks} in \GPTfour{} are in conjoined statements with other atomic checks, which may explain the relatively higher average completeness scores between the two models.

\begin{tcolorbox}[title=RQ1 summary: Qualitative analysis of Postconditions for \humanevalplus{}]
    \noindent We qualitatively identify nine atomic component categories of LLM-generated postconditions. 
    While we observe minimal correctness differences, bug completeness varied significantly; the weakest postcondition type, \texttt{Type Checks}, killed only 14\% of natural bugs on average while the strongest, \texttt{Arithmetic Equality} check, killed 82\%, a 6x difference. 
\end{tcolorbox}

\section{RQ2: Can \NLSpec{} help catch real world bugs?} 
\label{sec:rq2}

Beyond understanding whether LLMs can capture natural language intent via executable postconditions, we also want to understand \NLSpec{}'s real-world potential. To do so, we investigate the second motivating use case in Section~\ref{sec:motivation}: finding bugs in an existing code base. We evaluate \NLSpec{}'s bug-catching potential using \dfj{}~\cite{just2014defects4j}, a benchmark of historical Java bugs.

\subsection{RQ2--Research Methodology and Experimental Setup}

We outline our methodology for evaluating the capabilities of postconditions to catch real-world bugs: we describe the target benchmark \dfj{}, discuss prompt variations for Java, and provide our criteria for bug-discriminating postconditions. We model our approach after TOGA's approach~\cite{dinella2022toga}, where the goal is to find specifications/tests that a user could have used to catch a bug as they fail on the buggy version, and succeed on the fixed version. 

\subsubsection{Benchmark: \dfj{}}
\label{subsec:dfjBenchmark} 
For our experiments, 
we use \dfj{} 2.0~\cite{just2014defects4j}, a well-known benchmark of 835 manually curated real-world bugs gathered from 17 Java projects. For each bug, the dataset contains a set of bug-reproducing test cases (trigger tests), and regression test cases which load the class in which the method under test is contained. Each bug in \dfj{} also contains buggy and fixed versions of the code. We consider a postcondition to be test-set-correct if it passes all trigger and regression tests on the fixed version.

As our prompt leverages functional syntax introduced in Java 8 (see the postcondition in fig.~\ref{fig:widthPostcond} as an example), we only consider  the subset pf 525 bugs from \dfj{} that are reproducible when compiled targeting Java 8. Each bug may involve changes to multiple functions, for which we each generate postconditions. In total 840 functions are modified across the 525 bugs.

 \subsubsection{Bug Discriminating Postconditions}

\label{subsec:MethodsDfjPostconditionEvaluation}
To evaluate whether LLM-generated postconditions are capable of catching real-world bugs, we instrument the buggy and fixed function versions with each associated postcondition. We consider a generated postcondition to be \emph{bug-discriminating} if it satisfies the following criteria: 
\begin{enumerate}[leftmargin=0.21in]
    \item The postcondition \textbf{passes} all the trigger and regression tests, on the fixed version of a function.
    \item The postcondition \textbf{fails} a a trigger test or regression test on the buggy version of a function.
\end{enumerate}

The \dfj{} benchmark ensures that the difference between the buggy and fixed versions is minimized to only changes related to the bug-fix. Therefore, assuming a comprehensive test suite, any discriminating postcondition satisfying the above criteria is related to the (bug-related) change for the example. 
Finally, similar to our qualitative evaluation for RQ1 (see Section~\ref{subsec:humanEvalCorrectness}) we qualitatively analyze bug-discriminating postconditions to gain greater insight.

\subsubsection{Prompt Design and Ablations}
\label{subsec:MethodsDfjPostconditionGenderation}

To generate postcondtions for buggy functions in the dataset, we use the same prompt as in RQ1 (see fig.~\ref{fig:prompts}). Designed as language agnostic, the only change needed to adapt the prompt for \dfj{} is including additional code context. Given that \dfj{} problems are extracted from real-world projects, functions are comparatively more complex than those in \humanevalplus{} and are often tightly coupled with other project functions. Our initial investigations found that without some file-level context, LLMs rarely generate meaningful postconditions that also compile. Therefore, we include additional class and type-related context in the prompt. Given the limited context window of the LLMs used, we greedily include methods in the call graph for the buggy function (ordered by in-file placement) until the prompt tokens are exhausted. The call graph and in-file dependencies are determined using the Java language binding for \texttt{Tree-sitter}\footnote{\url{https://tree-sitter.github.io/tree-sitter/}}. 

For each buggy function, we combine of function and class-level in-file comments to formulate a natural language specification. In practice, this is primarily the buggy function's JavaDoc. We do not generate additional natural language (i.e., through code summarization) nor do we use external documentation: all natural language is pulled directly from the buggy function's source code file.

We choose to use only the \ic{simple} prompt from RQ1, as it led to more correct postconditions than did the \ic{base} version. Following the approach in RQ1, we report two variants of the prompt: 1) that only includes the natural language of the function 2) that includes both the natural language and the code of the buggy function body. 
\if False
\begin{enumerate}
    \item \ic{Simple} prompt with natural language and code context (no buggy code)
    \item \ic{Simple} prompt with  natural language, code context, and buggy implementation
\end{enumerate}
\fi
For each variant, we generate 10 postconditions for every function modified between the buggy and fixed projects (840 functions across 525 unique bugs). 

We choose to generate postconditions using two of the three earlier introduced models, used in RQ1: \GPTfour{} and \opensource{}. Given that \GPTfour{} and \GPTthree{} are comparable, closed-access chat models from OpenAI, we choose to focus on \GPTfour{} as it shows superior performance in RQ1. We choose to use \opensource{} as it is one of the few open-source chat-based models available. In total, we evaluate 33,600 postconditions (2 models \texttt{*} 2 ablations \texttt{*} 10 postconditions \texttt{*} \texttt{840} functions).

\subsection{\textit{RQ2}--Results: can LLM-generated postconditions catch real-world bugs?}

We detail our findings on if \NLSpec{} postconditions are test-set correct, and if they can catch bugs in real-world industrial-scale projects. We find that even with the increased complexity over \humanevalplus{}, \GPTfour{} is still able to produce correct postconditions for \dfj{} at a high rate. In addition, both \GPTfour{} and \opensource{} are able to generate bug-discriminating postconditions for a subset of \dfj{} bugs. All bug-discriminating postconditions were further analyzed via a qualitative analysis to gain insight into the ability of LLMs to catch bugs via \NLSpec{}. 

\subsubsection{Test-set correctness}
\begin{table}[t!]
\scriptsize
\begin{tabular}{lc|rrr|rrr|r}
\toprule
\multirow{3}{*}{Model} &
  \multicolumn{1}{l|}{\multirow{3}{*}{\begin{tabular}[c]{@{}l@{}}Prompt has:\\ NL Only = \color{deepred}\xmark\\ buggy code = \color{deepgreen}\checkmark\end{tabular}}} &
  \multicolumn{3}{c|}{\multirow{2}{*}{Compiles}} &
  \multicolumn{3}{c|}{\multirow{2}{*}{Test-set correct}} &
  \multicolumn{1}{r}{\multirow{3}{*}{\begin{tabular}[c]{@{}r@{}}\# disting- \\uishable\\ bugs \end{tabular}}} \\
 &
  \multicolumn{1}{l|}{} &
  \multicolumn{3}{c|}{} &
  \multicolumn{3}{c|}{} &
  \multicolumn{1}{l}{}  \\
 &
  \multicolumn{1}{l|}{} &
  @1 &
  @5 &
  @10 &
  @1 &
  @5 &
  @10 &
  \multicolumn{1}{l}{}  \\ \midrule

\GPTfour &
  \color{deepred}\xmark &
  0.65 &
  0.86 &
  0.89 &
  0.32 &
  0.57 &
  0.66 &
  35 
   \\
\GPTfour &
  \color{deepgreen}\checkmark &
  0.73 &
  0.90 &
  \textbf{0.93} &
  0.39 &
  0.66 &
  \textbf{0.75} &
  \textbf{47}
   \\ \midrule

\opensource &
  \color{deepred}\xmark    & 0.25
   & 0.68
   & 0.83
   & 0.11
   & 0.38
   & 0.55
   & 19
   \\
\opensource &
  \color{deepgreen}\checkmark  & 0.29
   & 0.72
   & \textbf{0.84}
   & 0.12
   & 0.39
   & \textbf{0.56}
   & \textbf{24}

   \\ \bottomrule
\end{tabular}
\caption{Table containing our \dfj{} results for postconditions generated for 840 methods across 525 historical bugs. We report the likelihood of generated postconditions to compile, and the likelihood that they pass all tests when instrumenting the fixed function (\textit{test-set correct} columns). \emph{ \# distinguishable bugs} is the number of bugs for which at least one generated postcondition was discriminating (see Section~\ref{subsec:MethodsDfjPostconditionEvaluation}).
\label{tab:defects4J}}
\vspace{-16pt}
\end{table}

Our full test-set correctness results for \dfj{} are in table~\ref{tab:defects4J}. We find that while lower than the results from \humanevalplus{}, \GPTfour{} still generate a significant number of test-set correct postconditions with respect to the fixed version of a function (e.g., correct with respect to programmer intent), achieving \texttt{accept@1} of up to 0.39 and \texttt{accept@10} of up to 0.75. \opensource{} performs worse, with \texttt{accept@1} and \texttt{accept@10} of 0.12 and 0.56 respectively. We note that these numbers may be higher in practice if postconditions are filtered by those that compile (see table~\ref{tab:defects4J}, \texttt{Compiles} column). In general, including the buggy code in the prompt leads to more test-set correct postconditions. This contrasts with the results from \humanevalplus{}, where we did not observe a difference. We hypothesize that this is the case because of (a) the comments not being completely descriptive, and (b)  the increased program and object complexity in \dfj{}, as supported by the fact that postconditions are also less likely to compile when the buggy code is omitted from the prompt.

\subsubsection{Bug-discriminating postconditions}

We find that LLMs can generate postconditions that distinguish between buggy and fixed code in real-world projects with respect to regression and trigger tests. As seen in Table~\ref{tab:defects4J}, \GPTfour{} was able to generate discriminating postconditions for up to 47/525 (9\%) bugs. \opensource{} caught fewer, but still generated postconditions that distinguished up to 25 bugs. Across all prompt variants and models, we were able to generate a bug-discriminating postcondition for 70 buggy methods from 64 unique bugs in \dfj{}, 12.2\% of all bugs considered.

\begin{tcolorbox}[title=RQ2 summary: Correctness and bug catching power on \dfj{}]    
\noindent We find that \NLSpec{} postconditions are often test-set correct for real-world functions (\texttt{accept@10} up to 0.75) and can be powerful enough to catch real-world bugs (\NLSpec{} discriminates 70 buggy methods from 64 bugs in \dfj{}). 
\end{tcolorbox}

\subsubsection{Qualitative analysis of bug-discriminating postconditions}

\label{subsec:dfjqualitative}
We conduct a qualitative evaluation of the bug-discriminating postconditions to gain insight into how \NLSpec{} postconditions discriminate real-world bugs. 
 We observed additional evidence both motivating the potential usefulness of \NLSpec{} and examples of why LLMs may be a good tool to solve this problem. To communicate these findings, we detail two cases.

\begin{figure}
\footnotesize
\begin{subfigure}{0.95\textwidth}
    \small
        \caption{Buggy function stub and javadoc.}
\begin{Java}
  /** Render the text and return the rendered Options in a StringBuffer.
  * @param width The number of characters to display per line
  * @param nextTab The position on the next line for the first tab.
  * @param text The text to be rendered.*/
  StringBuffer renderWrappedText(StringBuffer sb, int   width, int nextTab, String text);
\end{Java}
        \label{fig:enter-label}

    \end{subfigure}
    \begin{subfigure}{0.95\textwidth}
       
        \caption{ \label{fig:codeSober}Bug report indicating that the function sometimes erroneously renders text with more than \texttt{width} characters per line, behavior that directly conflicts with the Javadoc.}
        \setlength{\FrameSep}{5pt}
        \begin{framed}
        \footnotesize
         \color{deepred}
  \texttt{The method... has couple of bugs in the way that it deals with the [nextTab] variable. This causes it to format every line beyond the first line by [nextTab] too many characters \textbf{beyond the specified width}.}  

        \end{framed}
    \end{subfigure}
    
    \begin{subfigure}{0.95\textwidth}
        \caption{\label{fig:widthPostcond}Bug catching \NLSpec{} postcondition generated by \GPTfour{}. \texttt{rVal} is the function return value. This postcondition was generated without the buggy function code in the prompt.}
        \begin{Java}
 // Checks if the rendered text does not exceed the specified width per line
 assert rVal.toString().lines().allMatch(line -> line.length() <= width);
        \end{Java}
    \end{subfigure}

    \begin{subfigure}{0.95\textwidth}
        \caption{\label{fig:togaCliOracle}Bug-catching test from TOGA where TOGA expects this test prefix to through a \texttt{RuntimeException}. While this catches the bug, it is semantically removed from the bug's root cause.}
\begin{Java}
  public void test27()  throws Throwable  {
    HelpFormatter helpFormatter0 = new HelpFormatter();
    MockPrintWriter mockPrintWriter0 = new MockPrintWriter("-");
    helpFormatter0.printUsage((PrintWriter) mockPrintWriter0, 0, "[ Options: [ sh6ort "); }
\end{Java}
    \vspace{-15pt}
\end{subfigure}
    \caption{ 
    \label{fig:cliDfjExample} 
    Example from \dfj{} (Cli project, bug 8) where the bug can be caught via \NLSpec{}. Cli 8 is a bug in the implementation for calculating the width of lines when wrapping output text. The natural language function description specifically says that each line must be at most \texttt{width} characters long. \GPTfour{} translates this intent into the provided postcondition, which correctly catches the bug.
    }
        \vspace{-10pt}
\end{figure}

The first case is a historical bug from the Apache Commons CLI project.\footnote{Project page: \url{https://commons.apache.org/proper/commons-cli/}, Bug: \url{https://issues.apache.org/jira/browse/CLI-151}} As shown in fig.~\ref{fig:cliDfjExample}, the program should render multi-line text such that 1) white space padding is added at the beginning of every line after the first one and 2) that no line length exceeds a specified \texttt{width}. The requirement that each line should be \texttt{width} characters long is clearly specified in the Javadoc. However, the program sometimes incorrectly rendered lines longer than \texttt{width} due to a bug in the white space padding implementation. In our evaluation, \GPTfour{} generated multiple postconditions that catch this bug, including the example in \ref{fig:widthPostcond}. These bug-catching postconditions were generated by both prompt variations. This example evidences both that 1) informal natural language can meaningfully telegraph code bugs and 2) modern LLMs, such as \GPTfour{}, can sufficiently formalize natural language intent to capture the disagreement. Overall, this example shows the potential of \NLSpec{} to unearth coding inconsistencies solely from informal natural language documentation.

For our second example, we refer back to one of our initial examples motivating \NLSpec{} in Section~\ref{sec:motivation}, fig.~\ref{fig:mathDfjExample}. This example was adapted from \dfj{}, and consists of a historical bug in another popular Apache library project, Commons Math.\footnote{Project page: \url{https://commons.apache.org/proper/commons-math/}, Bug: \url{https://issues.apache.org/jira/browse/MATH-938}} In this bug, a method returning a reversed instance of a mathematical \texttt{Line} object does not retain sufficient precision in its internal state. \GPTfour{} is again able to generate multiple postconditions that correctly detect this bug: both postconditions in fig.~\ref{fig:mathDfjExample} are actual postconditions from our evaluation generated using the prompt with the buggy code included. As with the first example, this example demonstrates the potential of LLMs to generate postconditions powerful enough to capture real-world bugs. However, this example additionally provides evidence that LLMs in particular are helpful for realizing \NLSpec{}. Both postconditions detect the bug by leveraging general mathematical knowledge about the properties of a reversed line. The second postcondition in particular exemplifies the ability of LLMs to dynamically combine methods such as \texttt{dotProduct} from the project file's context with algebraic world knowledge that is external to the project's code.

\subsection{Baseline comparison: TOGA, Daikon}
To contextualise our results, we provide an empirical and qualitative comparison of the effectiveness of \NLSpec{} with respect to two other popular methods of inferring test oracles and invariant specifications. We choose a state-of-the-art technique for each: (a) TOGA~\cite{dinella2022toga}, a neural approach to generating test oracles, and (b) Daikon~\cite{ernst1999dynamically}, a popular technique to infer program invariants (including method postconditions) from multiple dynamic executions. 
There exists related efforts on generating unit tests neurally such as AthenaTest~\cite{tufano2021unit} but no public release exists for evaluating it for our setup.\footnote{Personal communication with the author of AthenaTest.} We focus our comparison on \dfj{} and the ability of each technique to generate correct tests or postconditions that distinguish historical bugs: to the best of our knowledge, neither TOGA nor Daikon support Python (and thus are not compatible with \humanevalplus{}). 

\subsubsection{TOGA}

TOGA is a neural approach to generating test oracles for a focal method. 
Given a test prefix, TOGA generates an assertion or expected exception that the test prefix is expected to satisfy. Although both approaches can generate assertions that may not agree with the implementation of a focal method, there are fundamental differences between test oracle generation (as in TOGA) and specification generation (as in \NLSpec{}). 
\NLSpec{} infers method postconditions that are expected to hold for all inputs. These can not only be checked during testing, but also at runtime on unseen inputs and trigger assertion failures instead of producing corrupted values; TOGA generated assertions can only be applied at testing time, since the assertions apply to the specific test prefix that reaches the buggy location. 
On the other hand, there may be some (algebraic) specifications that are best expressed over multiple method calls (e.g., {\tt s.pop(s.push(5)) == 5} for a stack object {\tt s} and specific values of inputs such as 5); expressing such a specification as a method postcondition (for either {\tt s.pop} or {\tt s.push} for even a single value 5) will require adding auxiliary ghost variables. 
Finally, assertions in test oracles are most often equalities (to match the expected output value on the specific input), whereas the assertions in method postconditions can be arbitrary Boolean expressions to capture all possible output values (see Table~\ref{tab:taxonomy} for some examples). 

\paragraph{\bf Setup.} We compare \NLSpec{}'s results on \dfj{} with the results reported on TOGA~\cite{dinella2022toga}. To enable TOGA to catch bugs without access to the failing trigger test, Dinella \emph{et al.} integrated TOGA with EvoSuite~\cite{fraser2011evosuite}, a popular automated testing tool. 
We used the set of 57 bugs found by TOGA (by reproducing their experimental setup released as a docker), each accompanied by a EvoSuite-generated test prefix and the corresponding test oracle. 
Of the 57 bugs reported by TOGA, 15 bugs were excluded from our \NLSpec{} evaluation due to Java limitations (Section~\ref{subsec:dfjBenchmark}).

\noindent\paragraph{\bf Evaluation Results.} Overall, we find that the bug finding capabilities of TOGA and \NLSpec{} are complementary. Of the 101 distinct bugs caught by at least one approach, only 5 are caught by both \NLSpec{} and TOGA.\footnote{The 5 in common are Cli 8, Cli 32, JacksonCore 8, Jsoup 88, and Math 99} 37 are only caught by TOGA, while 59 are only caught by \NLSpec{}. To better understand the differences between the two techniques, we conduct a qualitative evaluation of bugs caught by at least one approach. We note the following observations:

\begin{figure}
    \small
    \centering
    \begin{subfigure}{0.95\textwidth}
     \caption{\label{fig:compressToga}TOGA test oracle that catches Compress 11. TOGA finds the bug by simulating a small file and then explicitly catching the resulting exception.}
    \begin{Java}
 public void test16()  throws Throwable  {
    byte[] byteArray0 = new byte[179];
    ByteArrayInputStream inputStream0 = new ByteArrayInputStream(byteArray0);
    ArchiveStreamFactory archiveStreamFactory0 = new ArchiveStreamFactory();
    try { 
      archiveStreamFactory0.createArchiveInputStream((InputStream) inputStream0);
      fail("Expecting exception: Exception");
    } catch(Exception e) {  
      verifyException("org.apache.commons.compress.archivers.ArchiveStreamFactory", e);
    }}
    \end{Java}
    \end{subfigure} 

    \begin{subfigure}{0.95\textwidth}
    \caption{\label{fig:compressDaikon}Daikon postcondition distinguishes Compress 11. It does so as the buggy function involves a using a \texttt{ArchiveStream} factory function that can change the class name of the Input Stream class.  }
    \begin{Java}
 \old(in.getClass().getName()) == java.io.BufferedInputStream.class.getName()
    \end{Java} 
    \end{subfigure}
    \vspace{-10pt}
    \caption{TOGA test oracle and Daikon postcondition for a historical bug caught by both TOGA and Daikon, but not by \NLSpec{} (Compress 11). This bug involved incorrectly processing files less than 512 bytes files as tar archives, 
    and it was fixed by throwing an exception. 
    \label{fig:compress11}}
\end{figure}

\begin{itemize}[leftmargin=0.2in]
\item A majority of \NLSpec{}-caught bugs (52/64) could not be found by TOGA, due to the lack of any EvoSuite generated test prefix that reaches the bug location. This includes \textit{Math 9}, one of the two \dfj{} bugs we use to motivate \NLSpec{} (see fig.~\ref{fig:mathDfjExample}). This demonstrates the usefulness of \NLSpec{}'s ability to be checked at runtime on unseen inputs.

\item A majority of TOGA-caught bugs are ``exceptional'' bugs where the buggy code either throws an {\it unexpected exception} or fails to throw an {\it expected exception}. Since we do not currently model exceptional postconditions in \NLSpec{}, we fail to find most of these bugs. Fig.~\ref{fig:compress11} shows how leveraging a model for predicting exceptional postconditions enables TOGA to catch bugs that \NLSpec{} does not. 
Incorporating exceptional postconditions into \NLSpec{} is an intriguing direction for future work. Beyond exceptional postconditions, we also find that TOGA can model test prefixes that involve objects from different classes and methods (similar to the stack example with {\tt push} and {\tt pop}).

\item For the 5 common bugs, we observe that TOGA and \NLSpec{} find the 
same underlying bug with different means. For example, one of our motivating examples for \NLSpec{}, \textit{Cli 8}, is also caught by a TOGA test oracle. While both are helpful, \NLSpec{}'s assertion directly captures the semantics of the root cause of the bug (useful for both fault localization and patch construction). TOGA, however, provides a higher-level end-to-end test that is more removed from the buggy method, necessitating the developer spend additional time for root cause analysis. We present both bug catches for \textit{Cli 8} in fig.~\ref{fig:togaCliOracle}.  

\end{itemize}

\if False
\begin{figure}
\footnotesize
\begin{subfigure}{0.95\textwidth}
        \caption{Bug catching postcondition generated by \NLSpec{} which captures the semantic root cause of the bug. 
        }
        \begin{Java}
 // Checks if the rendered text does not exceed the specified width per line
 assert rVal.toString().lines().allMatch(line -> line.length() <= width);
        \end{Java}
\end{subfigure}

\begin{subfigure}{0.95\textwidth}
        \caption{\label{fig:togaCliOracle}Bug-catching test from TOGA where TOGA expects this test prefix to through a \texttt{RuntimeException}. While this catches the bug, it is semantically removed from the bug's root cause.}
\begin{Java}
  public void test27()  throws Throwable  {
    HelpFormatter helpFormatter0 = new HelpFormatter();
    MockPrintWriter mockPrintWriter0 = new MockPrintWriter("-");
    helpFormatter0.printUsage((PrintWriter) mockPrintWriter0, 0, "[ Options: [ sh6ort "); }
\end{Java}
\end{subfigure}

\caption{Comparison of \NLSpec{} and TOGA on \textit{Cli 8} (see Fig.~\ref{fig:cliDfjExample}). This bug causes lines in a text to be longer than expected, even after going through a line wrapping function, due to an issue with an increasing tab offset at the start of each line. \label{fig:cli8Comparison}}   
\end{figure}

\fi

\begin{figure}
\small
\begin{subfigure}{0.95\textwidth}
        \caption{Example bug-catching post conditions generated by \NLSpec{} which correctly asserts that the domain of a continuous distribution function should be greater than or equal zero. This postcondition catches a large number of bug-triggering inputs for this method. \label{fig:math95nl2spec}}
        \begin{Java}
  // Checks if the returnValue is greater than or equal to zero
  assert returnValue >= 0;
        \end{Java}
\end{subfigure}

\begin{subfigure}{0.95\textwidth}
    \caption{\label{fig:math95daikon}Daikon postcondition that distinguishes Math 9, but overfitts to the regression tests. }
    \begin{Java}
  daikon.Quant.fuzzy.eq(\result, 1.000020000400008) || daikon.Quant.fuzzy.eq(\result, 1.5)
    \end{Java} 
\end{subfigure}

\begin{subfigure}{0.95\textwidth}
\caption{ \label{fig:math95overview} Math 95 from \dfj{}: This function returns a domain for use by an Inverse Cumulative Probability function. The buggy version did not have sufficient bounds on  \texttt{getDenominatorDegreesOfFreedom}, leading to a potential negative domain (impossible for a cumulative Probability function) or a division by zero error. Highlighted tokens are those that were added for the fixed version.  }
\begin{Java}
    /** Access the initial domain value, based on <code>p</code>, used to
     * bracket a CDF root.  This method is used by
     * {@link #inverseCumulativeProbability(double)} to find critical values.
     * @param p the desired probability for the critical value
     * @return initial domain value */
    protected  double getInitialDomain (double p)  {
        double ret @= 1.0@ ;
        double d = getDenominatorDegreesOfFreedom();
        @if (d > 2.0) {@
            ret = d / (d - 2.0);
        @}@
        return ret; }
\end{Java}
\end{subfigure}
\caption{Comparison of \NLSpec{}, TOGA, and Daikon on \textit{Math 95}.  \label{fig:math95comparison}}   
\end{figure}

\subsubsection{Daikon} Daikon~\cite{ernst1999dynamically} uses multiple program runs to dynamically infer program invariants, including postconditions. Unlike \NLSpec{}, Daikon invariants are always implementation-consistent (it only retains expressions that are true across tested executions) and can only be generated from testable code (e.g., can not be generated from natural language alone).

\paragraph{\bf Setup.} We used Daikon to generate likely invariants from running the set of regression tests (without any failing trigger tests) on the buggy version. We then check if these specifications are bug-discriminating. We run Daikon using standard parameters for each buggy method to generate a set of likely postconditions. Due to challenges integrating Daikon with several of the projects in \dfj{}, we scope our evaluation to the 101 bugs found by either \NLSpec{} or TOGA.

\paragraph{\bf Results.} Overall, we find that while Daikon generates many  postconditions that are consistent with all tests, bug-discriminating postconditions are rare. 
Daikon generated postconditions for an associated buggy method for the majority of tested bugs (78/101). For the rest, Daikon either failed to generate any method postconditions on the buggy version using just the regression tests (17/101), or timed out after 10 minutes (6/101). 
The number of postconditions generated for any given method varied widely. 
However, we  observed only three instances of a Daikon-generated postcondition that is bug-discriminating.  
Daikon finds one bug that is not found by \NLSpec{}, but the other two specifications are incorrect. 
Fig.~\ref{fig:compressDaikon} shows the case where Daikon is able to catch a bug that \NLSpec{} does not, by detecting a class name change instigated through a factory function. 
For the remaining two cases, the bug-discriminating postcondtions overfit the regression tests and do not hold for all inputs. 
For example, the specification for Math 95 in Fig.~\ref{fig:math95daikon}) states that a return value should be close to one of two values $\{1.0, 1.5\}$. 
However, the fixed program admits many  more positive return values; this is correctly reflected by the \NLSpec{} postcondition in Fig.~\ref{fig:math95nl2spec}. 
In general, we observe that Daikon generated invariants are either very weak (e.g., a field is not modified), or are incorrect (do not generalize to all inputs). 
To the best of our understanding of \dfj{}, this is in contrast to the majority of bug-discriminating \NLSpec{} postconditions.

\begin{tcolorbox}[title=Baseline Comparison with TOGA and Daikon]    
\noindent Compared to two other approaches, we find that \NLSpec{} postconditions are either more widely applicable or find more bugs. We also note that the bugs found via \NLSpec{} are largely non-overlapping with those found by TOGA, indicating that the two approaches may be complementary.
\NLSpec{} finds many more bugs compared to Daikon, which usually generates invariants that overfit the observed executions.

\end{tcolorbox}

\section{Related Work}
\label{sec:related}

\paragraph{Specification Generation} A specification provides a comprehensive description of a program's intended behavior, encompassing the functional relationships between inputs and outputs, as well as the internal state dynamics. Specifications may vary in formality, ranging from informal descriptions such as API documentation to formal representations like test cases or runtime assertions. The applications of program specifications are extensive and include bug identification~\cite{jackson1992aspect, arcuri2008automation}, verification~\cite{chalin2006beyond, mike2004spec}, specification-driven development ~\cite{rutledge2014formal, ostroff2004agile, leitner2007contract}, code comprehension~\cite{bowen1993formal}. Our goal is to generate formal and functional specifications in the form of postconditions, articulating the desired input-output relationship of a code, given the informal natural language description. There has been a long line of work for automatically inferring specifications using static analysis~\cite{shoham2007static}, abstract interpretation~\cite{cousot2012abstract}, dynamic analysis~\cite{ernst1999dynamically}, and so on. While most of these existing works rely on a code implementation inferring the specification of existing code, our approach is to infer the desired behavior of the code from natural language. Similar to us, several approaches attempted to generate specification by analyzing API documentation or code comments using different natural language processing techniques such as named pattern matching~\cite{tan2012tcomment, pandita2012inferring, tan2007icomment, tan2011acomment}, text normalization~\cite{blasi2018translating}, entity recognition~\cite{zhong2009inferring}, natural language parsing~\cite{zhou2017analyzing}, etc. Being dependent on mostly hand-crafted rules and heuristics, most of these techniques only work on the semi-structured natural language format of the input and are not easily extensible across different programming languages and domains. In contrast, our technique relies on LLMs for world knowledge and our experiment shows the extensibility of our technique in two different languages -- Python and Java.

\paragraph{Machine Learning for Specifications} Machine Learning approaches for specification generation have shown promise in several orthogonal directions, including synthesizing test oracles~\cite{atlas_2022, mastropaolo2022using, dinella2022toga}, improving test coverage~\cite{lemieux2023codamosa}, generating unit tests~\cite{athenatest_2020, lahiri2022interactive} and so on. Depending on the application scenario, the specifications generated by these approaches are dependent on different inputs. 
AthenaTest~\cite{athenatest_2020} generates both the input and the oracle of a unit test from the implementation of the focal method (recall TOGA only generates test oracle). Closer to our work, TiCoder~\cite{lahiri2022interactive} leverages LLM to generate test input and output to formalize the user intent. While these approaches focus on generating concrete test cases (and potentially oracles), our approach is geared toward generating abstract functional relationships between the input and output of a procedure, which allows us to reason about a range of inputs.  
Similar to our work, EvoSpex~\cite{molina2021evospex} generates functional relationships of input-output with evolutionary learning. While their approach is aimed at summarizing existing program behavior (and therefore cannot be used to find bugs), our approach contributes towards generating formal specifications of desired input-output behavior. Recent work by \citet{vikram2023can} proposes to leverage LLM for generating property-based tests (PBTs). 
Speculyzer~\cite{key2022speak} uses LLMs to enumerate likely properties and inputs similar to PBT, but use them as heuristics to improve code generation.
Unlike our work, they do not seek to evaluate the correctness and completeness of these specifications.
In addition to the input-output specification generation, machine learning has been applied to generate intermediate specifications of a code such as invariants, using traditional machine learning~\cite{garg-popl16, sharma2016invariant}, deep learning~\cite{yao2020learning, ryancln2inv}, and LLMs~\cite{pei2023can}.

\section{Limitations and Threats to Validity}
\label{sec:threats}

\paragraph{LLM-Related Approach Limitations.} We note that there are several inherent weaknesses of our approach relating to the use of LLMs. In particular, we note that as we are using popular LLMs as a black box, the underlying model is not well understood. This can lead to a lack of interpretability of the results, as well as raise questions regarding result generalizability. In addition, due to the quickly evolving AI landscape, the results may become obsolete quickly. We consider some specific instances of these limitations in the rest of this section.

\emph{Data leakage.} One potential concern to the generalizability of the study is the use of popular benchmarks EvalPlus and
\dfj{} which are included in The Stack~\cite{kocetkov2022stack}, the dataset used to train StarChat, and may have been included as part of training datasets for both GPT-3.5 and GPT-4. The risk of data leakage could pose a threat to the internal validity of our study. Nevertheless, this concern is partially mitigated by the target task: the use of models to produce postconditions, which are not artifacts of either dataset. To our knowledge, postconditions have not been previously generated as part of any public-facing dataset. 

\emph{Stability of models' output.} Two of the models used in the experiments are accessed using OpenAI web APIs. OpenAI models are not open-access and are often updated or deprecated. This poses a threat to the replicability of our study. To mitigate this threat, we make available all postconditions generated by the closed-access models. We also use the open-access \opensource{}, and share all generated artifacts. In addition, we report results using the widely adopted metric \emph{accept@k}, which accounts for the stochasticity of model output.

\emph{Generalization of findings.}
Given the relatively small number of bugs (525) considered in the \dfj{} benchmark, our findings may not generalize to arbitrary bugs across different languages and repositories. We partially mitigate this threat by using real-world bugs from open-source projects and evaluating the capabilities of LLMs on both Python and Java benchmarks. In addition, the proposed taxonomy of postconditions (Section~\ref{sec:taxonomy}) is representative of only the programs in the \humanevalplus benchmark and may not generalize across languages or program complexities. 

\emph{Measure of postcondition completeness.} 
Our metric for postcondition completeness relies on a set of generated code mutants. The code mutants are generated to cover the space of possible bugs in the target function, however, the set of code mutants generated per problem will never represent a comprehensive set of possible bugs. Therefore, our measure of completeness is dependent on the range and quality of bugs covered in the set of mutants. This poses a threat to the internal validity of our study. To mitigate this threat we maximize the diversity of bugs by retaining only distinct mutants, and generate up to 233 buggy codes per problem.

\section{Conclusion}

In this paper, we introduce and define \NLSpec{} as the problem of translating natural language comments into programmatically checkable postconditions via LLMs. 
Our work proposes and validates metrics for assessing the correctness and completeness of postconditions derived from natural language, offering an initial step in systematizing the \NLSpec{} problem. Through an empirical and qualitative evaluation on two benchmarks, we find that LLMs are adept at translating natural language descriptions to formulate non-trivial postconditions that accurately capture programming intent. Our study also finds that LLM-generated postconditions can exhibit high discriminative power: we generate postconditions via \NLSpec{} that are able to discriminate 64 real-world historical bugs from industrial-scale Java projects. These findings underscore the feasibility and promise of leveraging natural language documentation into executable specifications. Our research highlights the possibility of LLMs acting as a bridge between informal language descriptions and formal code specifications, such that natural language comments can be used effectively to improve software validation and bug detection.
\section{Data Availability}

We plan to make a replication package publicly available in near future with our postcondition generation scripts and prompts, postcondition evaluation harness, and qualitative codebook. We also plan to make all generated postconditions available, along with the results of additional temperature ablations. Finally, we will include the unique natural and artificial LLM-generated mutants for \textit{EvalPlus}.


\bibliographystyle{ACM-Reference-Format}
\bibliography{main}


\begin{thebibliography}{58}


\ifx \showCODEN    \undefined \def \showCODEN     #1{\unskip}     \fi
\ifx \showDOI      \undefined \def \showDOI       #1{#1}\fi
\ifx \showISBNx    \undefined \def \showISBNx     #1{\unskip}     \fi
\ifx \showISBNxiii \undefined \def \showISBNxiii  #1{\unskip}     \fi
\ifx \showISSN     \undefined \def \showISSN      #1{\unskip}     \fi
\ifx \showLCCN     \undefined \def \showLCCN      #1{\unskip}     \fi
\ifx \shownote     \undefined \def \shownote      #1{#1}          \fi
\ifx \showarticletitle \undefined \def \showarticletitle #1{#1}   \fi
\ifx \showURL      \undefined \def \showURL       {\relax}        \fi
\providecommand\bibfield[2]{#2}
\providecommand\bibinfo[2]{#2}
\providecommand\natexlab[1]{#1}
\providecommand\showeprint[2][]{arXiv:#2}

\bibitem[Arcuri(2008)]%
        {arcuri2008automation}
\bibfield{author}{\bibinfo{person}{Andrea Arcuri}.}
  \bibinfo{year}{2008}\natexlab{}.
\newblock \showarticletitle{On the automation of fixing software bugs}. In
  \bibinfo{booktitle}{\emph{Companion of the 30th international conference on
  Software engineering}}. \bibinfo{pages}{1003--1006}.
\newblock


\bibitem[AWS(2023)]%
        {codewhisperer}
\bibfield{author}{\bibinfo{person}{Amazon AWS}.}
  \bibinfo{year}{2023}\natexlab{}.
\newblock \bibinfo{title}{Amazon CodeWhisperer}.
\newblock
\newblock
\newblock
\shownote{Accessed September 27, 2023.
  \url{https://aws.amazon.com/codewhisperer/}}.


\bibitem[Blasi et~al\mbox{.}(2018)]%
        {blasi2018translating}
\bibfield{author}{\bibinfo{person}{Arianna Blasi}, \bibinfo{person}{Alberto
  Goffi}, \bibinfo{person}{Konstantin Kuznetsov}, \bibinfo{person}{Alessandra
  Gorla}, \bibinfo{person}{Michael~D Ernst}, \bibinfo{person}{Mauro Pezz{\`e}},
  {and} \bibinfo{person}{Sergio~Delgado Castellanos}.}
  \bibinfo{year}{2018}\natexlab{}.
\newblock \showarticletitle{Translating code comments to procedure
  specifications}. In \bibinfo{booktitle}{\emph{Proceedings of the 27th ACM
  SIGSOFT international symposium on software testing and analysis}}.
  \bibinfo{pages}{242--253}.
\newblock


\bibitem[Bowen et~al\mbox{.}(1993)]%
        {bowen1993formal}
\bibfield{author}{\bibinfo{person}{Jonathan~P Bowen}, \bibinfo{person}{Peter~T
  Breuer}, {and} \bibinfo{person}{Kevin~C Lano}.}
  \bibinfo{year}{1993}\natexlab{}.
\newblock \showarticletitle{Formal specifications in software maintenance: From
  code to Z++ and back again}.
\newblock \bibinfo{journal}{\emph{Information and Software Technology}}
  \bibinfo{volume}{35}, \bibinfo{number}{11-12} (\bibinfo{year}{1993}),
  \bibinfo{pages}{679--690}.
\newblock


\bibitem[Chalin et~al\mbox{.}(2006)]%
        {chalin2006beyond}
\bibfield{author}{\bibinfo{person}{Patrice Chalin}, \bibinfo{person}{Joseph~R
  Kiniry}, \bibinfo{person}{Gary~T Leavens}, {and} \bibinfo{person}{Erik
  Poll}.} \bibinfo{year}{2006}\natexlab{}.
\newblock \showarticletitle{Beyond assertions: Advanced specification and
  verification with JML and ESC/Java2}. In \bibinfo{booktitle}{\emph{Formal
  Methods for Components and Objects: 4th International Symposium, FMCO 2005,
  Amsterdam, The Netherlands, November 1-4, 2005, Revised Lectures 4}}.
  Springer, \bibinfo{pages}{342--363}.
\newblock


\bibitem[Chen et~al\mbox{.}(2021)]%
        {chen2021evaluating}
\bibfield{author}{\bibinfo{person}{Mark Chen}, \bibinfo{person}{Jerry Tworek},
  \bibinfo{person}{Heewoo Jun}, \bibinfo{person}{Qiming Yuan},
  \bibinfo{person}{Henrique Ponde de~Oliveira Pinto}, \bibinfo{person}{Jared
  Kaplan}, \bibinfo{person}{Harri Edwards}, \bibinfo{person}{Yuri Burda},
  \bibinfo{person}{Nicholas Joseph}, \bibinfo{person}{Greg Brockman},
  {et~al\mbox{.}}} \bibinfo{year}{2021}\natexlab{}.
\newblock \showarticletitle{Evaluating large language models trained on code}.
\newblock \bibinfo{journal}{\emph{arXiv preprint arXiv:2107.03374}}
  (\bibinfo{year}{2021}).
\newblock


\bibitem[Cousot et~al\mbox{.}(2012)]%
        {cousot2012abstract}
\bibfield{author}{\bibinfo{person}{Patrick~M Cousot}, \bibinfo{person}{Radhia
  Cousot}, \bibinfo{person}{Francesco Logozzo}, {and} \bibinfo{person}{Michael
  Barnett}.} \bibinfo{year}{2012}\natexlab{}.
\newblock \showarticletitle{An abstract interpretation framework for
  refactoring with application to extract methods with contracts}. In
  \bibinfo{booktitle}{\emph{Proceedings of the ACM international conference on
  Object oriented programming systems languages and applications}}.
  \bibinfo{pages}{213--232}.
\newblock


\bibitem[Dijkstra and Scholten(1990)]%
        {dijkstra1990strongest}
\bibfield{author}{\bibinfo{person}{Edsger~W Dijkstra} {and}
  \bibinfo{person}{Carel~S Scholten}.} \bibinfo{year}{1990}\natexlab{}.
\newblock \showarticletitle{The strongest postcondition}.
\newblock \bibinfo{journal}{\emph{Predicate Calculus and Program Semantics}}
  (\bibinfo{year}{1990}), \bibinfo{pages}{209--215}.
\newblock


\bibitem[Dinella et~al\mbox{.}(2022)]%
        {dinella2022toga}
\bibfield{author}{\bibinfo{person}{Elizabeth Dinella}, \bibinfo{person}{Gabriel
  Ryan}, \bibinfo{person}{Todd Mytkowicz}, {and} \bibinfo{person}{Shuvendu~K
  Lahiri}.} \bibinfo{year}{2022}\natexlab{}.
\newblock \showarticletitle{Toga: A neural method for test oracle generation}.
  In \bibinfo{booktitle}{\emph{Proceedings of the 44th International Conference
  on Software Engineering}}. \bibinfo{pages}{2130--2141}.
\newblock


\bibitem[Ernst et~al\mbox{.}(1999)]%
        {ernst1999dynamically}
\bibfield{author}{\bibinfo{person}{Michael~D Ernst}, \bibinfo{person}{Jake
  Cockrell}, \bibinfo{person}{William~G Griswold}, {and} \bibinfo{person}{David
  Notkin}.} \bibinfo{year}{1999}\natexlab{}.
\newblock \showarticletitle{Dynamically discovering likely program invariants
  to support program evolution}. In \bibinfo{booktitle}{\emph{Proceedings of
  the 21st international conference on Software engineering}}.
  \bibinfo{pages}{213--224}.
\newblock


\bibitem[Fakhoury et~al\mbox{.}(2023)]%
        {fakhoury2023towards}
\bibfield{author}{\bibinfo{person}{Sarah Fakhoury}, \bibinfo{person}{Saikat
  Chakraborty}, \bibinfo{person}{Madan Musuvathi}, {and}
  \bibinfo{person}{Shuvendu~K Lahiri}.} \bibinfo{year}{2023}\natexlab{}.
\newblock \showarticletitle{Towards Generating Functionally Correct Code Edits
  from Natural Language Issue Descriptions}.
\newblock \bibinfo{journal}{\emph{arXiv preprint arXiv:2304.03816}}
  (\bibinfo{year}{2023}).
\newblock


\bibitem[Fraser and Arcuri(2011)]%
        {fraser2011evosuite}
\bibfield{author}{\bibinfo{person}{Gordon Fraser} {and} \bibinfo{person}{Andrea
  Arcuri}.} \bibinfo{year}{2011}\natexlab{}.
\newblock \showarticletitle{Evosuite: automatic test suite generation for
  object-oriented software}. In \bibinfo{booktitle}{\emph{Proceedings of the
  19th ACM SIGSOFT symposium and the 13th European conference on Foundations of
  software engineering}}. \bibinfo{pages}{416--419}.
\newblock


\bibitem[Garg et~al\mbox{.}(2016)]%
        {garg-popl16}
\bibfield{author}{\bibinfo{person}{Pranav Garg}, \bibinfo{person}{Daniel
  Neider}, \bibinfo{person}{P. Madhusudan}, {and} \bibinfo{person}{Dan Roth}.}
  \bibinfo{year}{2016}\natexlab{}.
\newblock \showarticletitle{Learning invariants using decision trees and
  implication counterexamples}. In \bibinfo{booktitle}{\emph{Proceedings of the
  43rd Annual {ACM} {SIGPLAN-SIGACT} Symposium on Principles of Programming
  Languages, {POPL} 2016, St. Petersburg, FL, USA, January 20 - 22, 2016}},
  \bibfield{editor}{\bibinfo{person}{Rastislav Bod{\'{\i}}k} {and}
  \bibinfo{person}{Rupak Majumdar}} (Eds.). \bibinfo{publisher}{{ACM}},
  \bibinfo{pages}{499--512}.
\newblock
\urldef\tempurl%
\url{https://doi.org/10.1145/2837614.2837664}
\showDOI{\tempurl}


\bibitem[GitHub(2023)]%
        {copilot}
\bibfield{author}{\bibinfo{person}{GitHub}.} \bibinfo{year}{2023}\natexlab{}.
\newblock \bibinfo{title}{GitHub Copilot}.
\newblock
\newblock
\newblock
\shownote{Accessed September 27, 2023.
  \url{https://github.com/features/copilot/}}.


\bibitem[Goffi et~al\mbox{.}(2016)]%
        {goffi2016automatic}
\bibfield{author}{\bibinfo{person}{Alberto Goffi}, \bibinfo{person}{Alessandra
  Gorla}, \bibinfo{person}{Michael~D Ernst}, {and} \bibinfo{person}{Mauro
  Pezz{\`e}}.} \bibinfo{year}{2016}\natexlab{}.
\newblock \showarticletitle{Automatic generation of oracles for exceptional
  behaviors}. In \bibinfo{booktitle}{\emph{Proceedings of the 25th
  international symposium on software testing and analysis}}.
  \bibinfo{pages}{213--224}.
\newblock


\bibitem[He(2019)]%
        {he2019understanding}
\bibfield{author}{\bibinfo{person}{Hao He}.} \bibinfo{year}{2019}\natexlab{}.
\newblock \showarticletitle{Understanding source code comments at large-scale}.
  In \bibinfo{booktitle}{\emph{Proceedings of the 2019 27th ACM Joint Meeting
  on European Software Engineering Conference and Symposium on the Foundations
  of Software Engineering}}. \bibinfo{pages}{1217--1219}.
\newblock


\bibitem[Hsu and Lachenbruch(2014)]%
        {hsu2014paired}
\bibfield{author}{\bibinfo{person}{Henry Hsu} {and} \bibinfo{person}{Peter~A
  Lachenbruch}.} \bibinfo{year}{2014}\natexlab{}.
\newblock \showarticletitle{Paired t test}.
\newblock \bibinfo{journal}{\emph{Wiley StatsRef: statistics reference online}}
  (\bibinfo{year}{2014}).
\newblock


\bibitem[Jackson(1992)]%
        {jackson1992aspect}
\bibfield{author}{\bibinfo{person}{Daniel Jackson}.}
  \bibinfo{year}{1992}\natexlab{}.
\newblock \emph{\bibinfo{title}{Aspect, a formal specification language for
  detecting bugs}}.
\newblock \bibinfo{thesistype}{Ph.\,D. Dissertation}.
  \bibinfo{school}{Citeseer}.
\newblock


\bibitem[Jia and Harman(2010)]%
        {jia2010analysis}
\bibfield{author}{\bibinfo{person}{Yue Jia} {and} \bibinfo{person}{Mark
  Harman}.} \bibinfo{year}{2010}\natexlab{}.
\newblock \showarticletitle{An analysis and survey of the development of
  mutation testing}.
\newblock \bibinfo{journal}{\emph{IEEE transactions on software engineering}}
  \bibinfo{volume}{37}, \bibinfo{number}{5} (\bibinfo{year}{2010}),
  \bibinfo{pages}{649--678}.
\newblock


\bibitem[Just et~al\mbox{.}(2014)]%
        {just2014defects4j}
\bibfield{author}{\bibinfo{person}{Ren{\'e} Just}, \bibinfo{person}{Darioush
  Jalali}, {and} \bibinfo{person}{Michael~D Ernst}.}
  \bibinfo{year}{2014}\natexlab{}.
\newblock \showarticletitle{Defects4J: A database of existing faults to enable
  controlled testing studies for Java programs}. In
  \bibinfo{booktitle}{\emph{Proceedings of the 2014 international symposium on
  software testing and analysis}}. \bibinfo{pages}{437--440}.
\newblock


\bibitem[Key et~al\mbox{.}(2022)]%
        {key2022speak}
\bibfield{author}{\bibinfo{person}{Darren Key}, \bibinfo{person}{Wen-Ding Li},
  {and} \bibinfo{person}{Kevin Ellis}.} \bibinfo{year}{2022}\natexlab{}.
\newblock \showarticletitle{I speak, you verify: Toward trustworthy neural
  program synthesis}.
\newblock \bibinfo{journal}{\emph{arXiv preprint arXiv:2210.00848}}
  (\bibinfo{year}{2022}).
\newblock


\bibitem[Kocetkov et~al\mbox{.}(2022)]%
        {kocetkov2022stack}
\bibfield{author}{\bibinfo{person}{Denis Kocetkov}, \bibinfo{person}{Raymond
  Li}, \bibinfo{person}{Loubna~Ben Allal}, \bibinfo{person}{Jia Li},
  \bibinfo{person}{Chenghao Mou}, \bibinfo{person}{Carlos~Mu{\~n}oz Ferrandis},
  \bibinfo{person}{Yacine Jernite}, \bibinfo{person}{Margaret Mitchell},
  \bibinfo{person}{Sean Hughes}, \bibinfo{person}{Thomas Wolf},
  {et~al\mbox{.}}} \bibinfo{year}{2022}\natexlab{}.
\newblock \showarticletitle{The stack: 3 tb of permissively licensed source
  code}.
\newblock \bibinfo{journal}{\emph{arXiv preprint arXiv:2211.15533}}
  (\bibinfo{year}{2022}).
\newblock


\bibitem[Lahiri et~al\mbox{.}(2022)]%
        {lahiri2022interactive}
\bibfield{author}{\bibinfo{person}{Shuvendu~K Lahiri}, \bibinfo{person}{Aaditya
  Naik}, \bibinfo{person}{Georgios Sakkas}, \bibinfo{person}{Piali Choudhury},
  \bibinfo{person}{Curtis von Veh}, \bibinfo{person}{Madanlal Musuvathi},
  \bibinfo{person}{Jeevana~Priya Inala}, \bibinfo{person}{Chenglong Wang},
  {and} \bibinfo{person}{Jianfeng Gao}.} \bibinfo{year}{2022}\natexlab{}.
\newblock \showarticletitle{Interactive code generation via test-driven
  user-intent formalization}.
\newblock \bibinfo{journal}{\emph{arXiv preprint arXiv:2208.05950}}
  (\bibinfo{year}{2022}).
\newblock


\bibitem[Lampinen et~al\mbox{.}(2022)]%
        {lampinen2022language}
\bibfield{author}{\bibinfo{person}{Andrew~K. Lampinen}, \bibinfo{person}{Ishita
  Dasgupta}, \bibinfo{person}{Stephanie C.~Y. Chan}, \bibinfo{person}{Kory
  Matthewson}, \bibinfo{person}{Michael~Henry Tessler},
  \bibinfo{person}{Antonia Creswell}, \bibinfo{person}{James~L. McClelland},
  \bibinfo{person}{Jane~X. Wang}, {and} \bibinfo{person}{Felix Hill}.}
  \bibinfo{year}{2022}\natexlab{}.
\newblock \bibinfo{title}{Can language models learn from explanations in
  context?}
\newblock
\newblock
\showeprint[arxiv]{2204.02329}~[cs.CL]


\bibitem[Leino(2010)]%
        {leino2010dafny}
\bibfield{author}{\bibinfo{person}{K~Rustan~M Leino}.}
  \bibinfo{year}{2010}\natexlab{}.
\newblock \showarticletitle{Dafny: An automatic program verifier for functional
  correctness}. In \bibinfo{booktitle}{\emph{Logic for Programming, Artificial
  Intelligence, and Reasoning: 16th International Conference, LPAR-16, Dakar,
  Senegal, April 25--May 1, 2010, Revised Selected Papers 16}}. Springer,
  \bibinfo{pages}{348--370}.
\newblock


\bibitem[Leitner et~al\mbox{.}(2007)]%
        {leitner2007contract}
\bibfield{author}{\bibinfo{person}{Andreas Leitner}, \bibinfo{person}{Ilinca
  Ciupa}, \bibinfo{person}{Manuel Oriol}, \bibinfo{person}{Bertrand Meyer},
  {and} \bibinfo{person}{Arno Fiva}.} \bibinfo{year}{2007}\natexlab{}.
\newblock \showarticletitle{Contract driven development= test driven
  development-writing test cases}. In \bibinfo{booktitle}{\emph{Proceedings of
  the the 6th joint meeting of the European software engineering conference and
  the ACM SIGSOFT symposium on The foundations of software engineering}}.
  \bibinfo{pages}{425--434}.
\newblock


\bibitem[Lemieux et~al\mbox{.}(2023)]%
        {lemieux2023codamosa}
\bibfield{author}{\bibinfo{person}{Caroline Lemieux},
  \bibinfo{person}{Jeevana~Priya Inala}, \bibinfo{person}{Shuvendu~K Lahiri},
  {and} \bibinfo{person}{Siddhartha Sen}.} \bibinfo{year}{2023}\natexlab{}.
\newblock \showarticletitle{CODAMOSA: Escaping coverage plateaus in test
  generation with pre-trained large language models}. In
  \bibinfo{booktitle}{\emph{International conference on software engineering
  (ICSE)}}.
\newblock


\bibitem[Li et~al\mbox{.}(2023)]%
        {li2023starcoder}
\bibfield{author}{\bibinfo{person}{Raymond Li}, \bibinfo{person}{Loubna~Ben
  Allal}, \bibinfo{person}{Yangtian Zi}, \bibinfo{person}{Niklas Muennighoff},
  \bibinfo{person}{Denis Kocetkov}, \bibinfo{person}{Chenghao Mou},
  \bibinfo{person}{Marc Marone}, \bibinfo{person}{Christopher Akiki},
  \bibinfo{person}{Jia Li}, \bibinfo{person}{Jenny Chim}, {et~al\mbox{.}}}
  \bibinfo{year}{2023}\natexlab{}.
\newblock \showarticletitle{StarCoder: may the source be with you!}
\newblock \bibinfo{journal}{\emph{arXiv preprint arXiv:2305.06161}}
  (\bibinfo{year}{2023}).
\newblock


\bibitem[Liu et~al\mbox{.}(2023)]%
        {liu2023IsYourCode}
\bibfield{author}{\bibinfo{person}{Jiawei Liu}, \bibinfo{person}{Chunqiu~Steven
  Xia}, \bibinfo{person}{Yuyao Wang}, {and} \bibinfo{person}{Lingming Zhang}.}
  \bibinfo{year}{2023}\natexlab{}.
\newblock \showarticletitle{{Is Your Code Generated by ChatGPT Really Correct?
  Rigorous Evaluation of Large Language Models for Code Generation}}.
\newblock \bibinfo{journal}{\emph{37th cconference on Neural Information
  processing Systems (NeurIPS), 2023}} (\bibinfo{year}{2023}).
\newblock
\urldef\tempurl%
\url{https://arxiv.org/abs/2305.01210}
\showURL{%
\tempurl}


\bibitem[Mastropaolo et~al\mbox{.}(2022)]%
        {mastropaolo2022using}
\bibfield{author}{\bibinfo{person}{Antonio Mastropaolo},
  \bibinfo{person}{Nathan Cooper}, \bibinfo{person}{David~Nader Palacio},
  \bibinfo{person}{Simone Scalabrino}, \bibinfo{person}{Denys Poshyvanyk},
  \bibinfo{person}{Rocco Oliveto}, {and} \bibinfo{person}{Gabriele Bavota}.}
  \bibinfo{year}{2022}\natexlab{}.
\newblock \showarticletitle{Using transfer learning for code-related tasks}.
\newblock \bibinfo{journal}{\emph{IEEE Transactions on Software Engineering}}
  \bibinfo{volume}{49}, \bibinfo{number}{4} (\bibinfo{year}{2022}),
  \bibinfo{pages}{1580--1598}.
\newblock


\bibitem[Mike et~al\mbox{.}(2004)]%
        {mike2004spec}
\bibfield{author}{\bibinfo{person}{B Mike}, \bibinfo{person}{K~Rustan~M Leino},
  {and} \bibinfo{person}{S Wolfram}.} \bibinfo{year}{2004}\natexlab{}.
\newblock \bibinfo{title}{The Spec\# programming system: An overview. In
  Construction and Analysis of Safe, Secure, and Interoperable Smart devices
  (CASSIS)” volume 3362 of Lecture Notes in Computer Science}.
\newblock
\newblock


\bibitem[Molina et~al\mbox{.}(2021)]%
        {molina2021evospex}
\bibfield{author}{\bibinfo{person}{Facundo Molina}, \bibinfo{person}{Pablo
  Ponzio}, \bibinfo{person}{Nazareno Aguirre}, {and} \bibinfo{person}{Marcelo
  Frias}.} \bibinfo{year}{2021}\natexlab{}.
\newblock \showarticletitle{EvoSpex: An evolutionary algorithm for learning
  postconditions}. In \bibinfo{booktitle}{\emph{2021 IEEE/ACM 43rd
  International Conference on Software Engineering (ICSE)}}. IEEE,
  \bibinfo{pages}{1223--1235}.
\newblock


\bibitem[Nijkamp et~al\mbox{.}(2022)]%
        {nijkamp2022codegen}
\bibfield{author}{\bibinfo{person}{Erik Nijkamp}, \bibinfo{person}{Bo Pang},
  \bibinfo{person}{Hiroaki Hayashi}, \bibinfo{person}{Lifu Tu},
  \bibinfo{person}{Huan Wang}, \bibinfo{person}{Yingbo Zhou},
  \bibinfo{person}{Silvio Savarese}, {and} \bibinfo{person}{Caiming Xiong}.}
  \bibinfo{year}{2022}\natexlab{}.
\newblock \showarticletitle{Codegen: An open large language model for code with
  multi-turn program synthesis}.
\newblock \bibinfo{journal}{\emph{arXiv preprint arXiv:2203.13474}}
  (\bibinfo{year}{2022}).
\newblock


\bibitem[Olausson et~al\mbox{.}(2023)]%
        {olausson2023demystifying}
\bibfield{author}{\bibinfo{person}{Theo~X Olausson},
  \bibinfo{person}{Jeevana~Priya Inala}, \bibinfo{person}{Chenglong Wang},
  \bibinfo{person}{Jianfeng Gao}, {and} \bibinfo{person}{Armando
  Solar-Lezama}.} \bibinfo{year}{2023}\natexlab{}.
\newblock \showarticletitle{Demystifying GPT Self-Repair for Code Generation}.
\newblock \bibinfo{journal}{\emph{arXiv preprint arXiv:2306.09896}}
  (\bibinfo{year}{2023}).
\newblock


\bibitem[Ostroff et~al\mbox{.}(2004)]%
        {ostroff2004agile}
\bibfield{author}{\bibinfo{person}{Jonathan~S Ostroff}, \bibinfo{person}{David
  Makalsky}, {and} \bibinfo{person}{Richard~F Paige}.}
  \bibinfo{year}{2004}\natexlab{}.
\newblock \showarticletitle{Agile specification-driven development}. In
  \bibinfo{booktitle}{\emph{International Conference on Extreme Programming and
  Agile Processes in Software Engineering}}. Springer,
  \bibinfo{pages}{104--112}.
\newblock


\bibitem[Ouyang et~al\mbox{.}(2022)]%
        {ouyang2022training}
\bibfield{author}{\bibinfo{person}{Long Ouyang}, \bibinfo{person}{Jeffrey Wu},
  \bibinfo{person}{Xu Jiang}, \bibinfo{person}{Diogo Almeida},
  \bibinfo{person}{Carroll Wainwright}, \bibinfo{person}{Pamela Mishkin},
  \bibinfo{person}{Chong Zhang}, \bibinfo{person}{Sandhini Agarwal},
  \bibinfo{person}{Katarina Slama}, \bibinfo{person}{Alex Ray},
  {et~al\mbox{.}}} \bibinfo{year}{2022}\natexlab{}.
\newblock \showarticletitle{Training language models to follow instructions
  with human feedback}.
\newblock \bibinfo{journal}{\emph{Advances in Neural Information Processing
  Systems}}  \bibinfo{volume}{35} (\bibinfo{year}{2022}),
  \bibinfo{pages}{27730--27744}.
\newblock


\bibitem[Pandita et~al\mbox{.}(2012)]%
        {pandita2012inferring}
\bibfield{author}{\bibinfo{person}{Rahul Pandita}, \bibinfo{person}{Xusheng
  Xiao}, \bibinfo{person}{Hao Zhong}, \bibinfo{person}{Tao Xie},
  \bibinfo{person}{Stephen Oney}, {and} \bibinfo{person}{Amit Paradkar}.}
  \bibinfo{year}{2012}\natexlab{}.
\newblock \showarticletitle{Inferring method specifications from natural
  language API descriptions}. In \bibinfo{booktitle}{\emph{2012 34th
  international conference on software engineering (ICSE)}}. IEEE,
  \bibinfo{pages}{815--825}.
\newblock


\bibitem[Pei et~al\mbox{.}(2023)]%
        {pei2023can}
\bibfield{author}{\bibinfo{person}{Kexin Pei}, \bibinfo{person}{David Bieber},
  \bibinfo{person}{Kensen Shi}, \bibinfo{person}{Charles Sutton}, {and}
  \bibinfo{person}{Pengcheng Yin}.} \bibinfo{year}{2023}\natexlab{}.
\newblock \showarticletitle{Can Large Language Models Reason about Program
  Invariants?}
\newblock  (\bibinfo{year}{2023}).
\newblock


\bibitem[Pfeiffer(2020)]%
        {pfeiffer2020constitutes}
\bibfield{author}{\bibinfo{person}{Rolf-Helge Pfeiffer}.}
  \bibinfo{year}{2020}\natexlab{}.
\newblock \showarticletitle{What constitutes software? An empirical,
  descriptive study of artifacts}. In \bibinfo{booktitle}{\emph{Proceedings of
  the 17th International Conference on Mining Software Repositories}}.
  \bibinfo{pages}{481--491}.
\newblock


\bibitem[Rutledge et~al\mbox{.}(2014)]%
        {rutledge2014formal}
\bibfield{author}{\bibinfo{person}{Richard Rutledge}, \bibinfo{person}{Sheryl
  Duggins}, \bibinfo{person}{Dan Lo}, {and} \bibinfo{person}{Frank Tsui}.}
  \bibinfo{year}{2014}\natexlab{}.
\newblock \showarticletitle{Formal specification-driven development}. In
  \bibinfo{booktitle}{\emph{Proceedings of the International Conference on
  Software Engineering Research and Practice (SERP)}}. The Steering Committee
  of The World Congress in Computer Science, Computer~…, \bibinfo{pages}{1}.
\newblock


\bibitem[Ryan et~al\mbox{.}(2020)]%
        {ryancln2inv}
\bibfield{author}{\bibinfo{person}{Gabriel Ryan}, \bibinfo{person}{Justin
  Wong}, \bibinfo{person}{Jianan Yao}, \bibinfo{person}{Ronghui Gu}, {and}
  \bibinfo{person}{Suman Jana}.} \bibinfo{year}{2020}\natexlab{}.
\newblock \showarticletitle{CLN2INV: Learning Loop Invariants with Continuous
  Logic Networks}. In \bibinfo{booktitle}{\emph{International Conference on
  Learning Representations}}.
\newblock


\bibitem[Sharma and Aiken(2016)]%
        {sharma2016invariant}
\bibfield{author}{\bibinfo{person}{Rahul Sharma} {and} \bibinfo{person}{Alex
  Aiken}.} \bibinfo{year}{2016}\natexlab{}.
\newblock \showarticletitle{From invariant checking to invariant inference
  using randomized search}.
\newblock \bibinfo{journal}{\emph{Formal Methods in System Design}}
  \bibinfo{volume}{48} (\bibinfo{year}{2016}), \bibinfo{pages}{235--256}.
\newblock


\bibitem[Shoham et~al\mbox{.}(2007)]%
        {shoham2007static}
\bibfield{author}{\bibinfo{person}{Sharon Shoham}, \bibinfo{person}{Eran
  Yahav}, \bibinfo{person}{Stephen Fink}, {and} \bibinfo{person}{Marco
  Pistoia}.} \bibinfo{year}{2007}\natexlab{}.
\newblock \showarticletitle{Static specification mining using automata-based
  abstractions}. In \bibinfo{booktitle}{\emph{Proceedings of the 2007
  International Symposium on Software Testing and Analysis}}.
  \bibinfo{pages}{174--184}.
\newblock


\bibitem[Swamy et~al\mbox{.}(2011)]%
        {swamy2011fstar}
\bibfield{author}{\bibinfo{person}{Nikhil Swamy}, \bibinfo{person}{Juan Chen},
  \bibinfo{person}{C\'{e}dric Fournet}, \bibinfo{person}{Pierre-Yves Strub},
  \bibinfo{person}{Karthikeyan Bhargavan}, {and} \bibinfo{person}{Jean Yang}.}
  \bibinfo{year}{2011}\natexlab{}.
\newblock \showarticletitle{Secure Distributed Programming with Value-Dependent
  Types}. In \bibinfo{booktitle}{\emph{Proceedings of the 16th ACM SIGPLAN
  International Conference on Functional Programming}} (Tokyo, Japan)
  \emph{(\bibinfo{series}{ICFP '11})}. \bibinfo{publisher}{Association for
  Computing Machinery}, \bibinfo{address}{New York, NY, USA},
  \bibinfo{pages}{266–278}.
\newblock
\showISBNx{9781450308656}
\urldef\tempurl%
\url{https://doi.org/10.1145/2034773.2034811}
\showDOI{\tempurl}


\bibitem[Tabnine(2023)]%
        {tabnine}
\bibfield{author}{\bibinfo{person}{Tabnine}.} \bibinfo{year}{2023}\natexlab{}.
\newblock \bibinfo{title}{Tabnine Code Completion}.
\newblock
\newblock
\newblock
\shownote{Accessed September 27, 2023. \url{https://www.tabnine.com/}}.


\bibitem[Tan et~al\mbox{.}(2007)]%
        {tan2007icomment}
\bibfield{author}{\bibinfo{person}{Lin Tan}, \bibinfo{person}{Ding Yuan},
  \bibinfo{person}{Gopal Krishna}, {and} \bibinfo{person}{Yuanyuan Zhou}.}
  \bibinfo{year}{2007}\natexlab{}.
\newblock \showarticletitle{/* iComment: Bugs or bad comments?*}. In
  \bibinfo{booktitle}{\emph{Proceedings of twenty-first ACM SIGOPS symposium on
  Operating systems principles}}. \bibinfo{pages}{145--158}.
\newblock


\bibitem[Tan et~al\mbox{.}(2011)]%
        {tan2011acomment}
\bibfield{author}{\bibinfo{person}{Lin Tan}, \bibinfo{person}{Yuanyuan Zhou},
  {and} \bibinfo{person}{Yoann Padioleau}.} \bibinfo{year}{2011}\natexlab{}.
\newblock \showarticletitle{aComment: mining annotations from comments and code
  to detect interrupt related concurrency bugs}. In
  \bibinfo{booktitle}{\emph{Proceedings of the 33rd international conference on
  software engineering}}. \bibinfo{pages}{11--20}.
\newblock


\bibitem[Tan et~al\mbox{.}(2012)]%
        {tan2012tcomment}
\bibfield{author}{\bibinfo{person}{Shin~Hwei Tan}, \bibinfo{person}{Darko
  Marinov}, \bibinfo{person}{Lin Tan}, {and} \bibinfo{person}{Gary~T Leavens}.}
  \bibinfo{year}{2012}\natexlab{}.
\newblock \showarticletitle{@ tcomment: Testing javadoc comments to detect
  comment-code inconsistencies}. In \bibinfo{booktitle}{\emph{2012 IEEE Fifth
  International Conference on Software Testing, Verification and Validation}}.
  IEEE, \bibinfo{pages}{260--269}.
\newblock


\bibitem[Tufano et~al\mbox{.}(2020)]%
        {athenatest_2020}
\bibfield{author}{\bibinfo{person}{Michele Tufano}, \bibinfo{person}{Dawn
  Drain}, \bibinfo{person}{Alexey Svyatkovskiy}, \bibinfo{person}{Shao~Kun
  Deng}, {and} \bibinfo{person}{Neel Sundaresan}.}
  \bibinfo{year}{2020}\natexlab{}.
\newblock \bibinfo{title}{Unit Test Case Generation with Transformers and Focal
  Context}.
\newblock
\newblock
\urldef\tempurl%
\url{https://doi.org/10.48550/ARXIV.2009.05617}
\showDOI{\tempurl}


\bibitem[Tufano et~al\mbox{.}(2021)]%
        {tufano2021unit}
\bibfield{author}{\bibinfo{person}{Michele Tufano}, \bibinfo{person}{Dawn
  Drain}, \bibinfo{person}{Alexey Svyatkovskiy}, \bibinfo{person}{Shao~Kun
  Deng}, {and} \bibinfo{person}{Neel Sundaresan}.}
  \bibinfo{year}{2021}\natexlab{}.
\newblock \bibinfo{title}{Unit Test Case Generation with Transformers and Focal
  Context}.
\newblock
\newblock
\showeprint[arxiv]{2009.05617}~[cs.SE]


\bibitem[Tufano et~al\mbox{.}(2022)]%
        {atlas_2022}
\bibfield{author}{\bibinfo{person}{Michele Tufano}, \bibinfo{person}{Dawn
  Drain}, \bibinfo{person}{Alexey Svyatkovskiy}, {and} \bibinfo{person}{Neel
  Sundaresan}.} \bibinfo{year}{2022}\natexlab{}.
\newblock \showarticletitle{Generating Accurate Assert Statements for Unit Test
  Cases using Pretrained Transformers}. In \bibinfo{booktitle}{\emph{{IEEE/ACM}
  International Conference on Automation of Software Test, AST@ICSE 2022,
  Pittsburgh, PA, USA, May 21-22, 2022}}. \bibinfo{publisher}{{ACM/IEEE}},
  \bibinfo{pages}{54--64}.
\newblock
\urldef\tempurl%
\url{https://doi.org/10.1145/3524481.3527220}
\showDOI{\tempurl}


\bibitem[Vikram et~al\mbox{.}(2023)]%
        {vikram2023can}
\bibfield{author}{\bibinfo{person}{Vasudev Vikram}, \bibinfo{person}{Caroline
  Lemieux}, {and} \bibinfo{person}{Rohan Padhye}.}
  \bibinfo{year}{2023}\natexlab{}.
\newblock \showarticletitle{Can Large Language Models Write Good Property-Based
  Tests?}
\newblock \bibinfo{journal}{\emph{arXiv preprint arXiv:2307.04346}}
  (\bibinfo{year}{2023}).
\newblock


\bibitem[Wei et~al\mbox{.}(2022)]%
        {wei2022emergent}
\bibfield{author}{\bibinfo{person}{Jason Wei}, \bibinfo{person}{Yi Tay},
  \bibinfo{person}{Rishi Bommasani}, \bibinfo{person}{Colin Raffel},
  \bibinfo{person}{Barret Zoph}, \bibinfo{person}{Sebastian Borgeaud},
  \bibinfo{person}{Dani Yogatama}, \bibinfo{person}{Maarten Bosma},
  \bibinfo{person}{Denny Zhou}, \bibinfo{person}{Donald Metzler},
  {et~al\mbox{.}}} \bibinfo{year}{2022}\natexlab{}.
\newblock \showarticletitle{Emergent Abilities of Large Language Models}.
\newblock \bibinfo{journal}{\emph{Transactions on Machine Learning Research}}
  (\bibinfo{year}{2022}).
\newblock


\bibitem[Wei et~al\mbox{.}(2023)]%
        {wei2023chainofthought}
\bibfield{author}{\bibinfo{person}{Jason Wei}, \bibinfo{person}{Xuezhi Wang},
  \bibinfo{person}{Dale Schuurmans}, \bibinfo{person}{Maarten Bosma},
  \bibinfo{person}{Brian Ichter}, \bibinfo{person}{Fei Xia},
  \bibinfo{person}{Ed Chi}, \bibinfo{person}{Quoc Le}, {and}
  \bibinfo{person}{Denny Zhou}.} \bibinfo{year}{2023}\natexlab{}.
\newblock \bibinfo{title}{Chain-of-Thought Prompting Elicits Reasoning in Large
  Language Models}.
\newblock
\newblock
\showeprint[arxiv]{2201.11903}~[cs.CL]


\bibitem[Yao et~al\mbox{.}(2020)]%
        {yao2020learning}
\bibfield{author}{\bibinfo{person}{Jianan Yao}, \bibinfo{person}{Gabriel Ryan},
  \bibinfo{person}{Justin Wong}, \bibinfo{person}{Suman Jana}, {and}
  \bibinfo{person}{Ronghui Gu}.} \bibinfo{year}{2020}\natexlab{}.
\newblock \showarticletitle{Learning nonlinear loop invariants with gated
  continuous logic networks}. In \bibinfo{booktitle}{\emph{Proceedings of the
  41st ACM SIGPLAN Conference on Programming Language Design and
  Implementation}}. \bibinfo{pages}{106--120}.
\newblock


\bibitem[Zhong et~al\mbox{.}(2009)]%
        {zhong2009inferring}
\bibfield{author}{\bibinfo{person}{Hao Zhong}, \bibinfo{person}{Lu Zhang},
  \bibinfo{person}{Tao Xie}, {and} \bibinfo{person}{Hong Mei}.}
  \bibinfo{year}{2009}\natexlab{}.
\newblock \showarticletitle{Inferring resource specifications from natural
  language API documentation}. In \bibinfo{booktitle}{\emph{2009 IEEE/ACM
  International Conference on Automated Software Engineering}}. IEEE,
  \bibinfo{pages}{307--318}.
\newblock


\bibitem[Zhou et~al\mbox{.}(2023)]%
        {zhou2023leasttomost}
\bibfield{author}{\bibinfo{person}{Denny Zhou}, \bibinfo{person}{Nathanael
  Schärli}, \bibinfo{person}{Le Hou}, \bibinfo{person}{Jason Wei},
  \bibinfo{person}{Nathan Scales}, \bibinfo{person}{Xuezhi Wang},
  \bibinfo{person}{Dale Schuurmans}, \bibinfo{person}{Claire Cui},
  \bibinfo{person}{Olivier Bousquet}, \bibinfo{person}{Quoc Le}, {and}
  \bibinfo{person}{Ed Chi}.} \bibinfo{year}{2023}\natexlab{}.
\newblock \bibinfo{title}{Least-to-Most Prompting Enables Complex Reasoning in
  Large Language Models}.
\newblock
\newblock
\showeprint[arxiv]{2205.10625}~[cs.AI]


\bibitem[Zhou et~al\mbox{.}(2017)]%
        {zhou2017analyzing}
\bibfield{author}{\bibinfo{person}{Yu Zhou}, \bibinfo{person}{Ruihang Gu},
  \bibinfo{person}{Taolue Chen}, \bibinfo{person}{Zhiqiu Huang},
  \bibinfo{person}{Sebastiano Panichella}, {and} \bibinfo{person}{Harald
  Gall}.} \bibinfo{year}{2017}\natexlab{}.
\newblock \showarticletitle{Analyzing APIs documentation and code to detect
  directive defects}. In \bibinfo{booktitle}{\emph{2017 IEEE/ACM 39th
  International Conference on Software Engineering (ICSE)}}. IEEE,
  \bibinfo{pages}{27--37}.
\newblock


\end{thebibliography}
\appendix

\end{document}